\documentclass[preprint,12pt]{elsarticle}
\biboptions{sort&compress}

\usepackage{euscript,color} 

\usepackage[T1]{fontenc}
\usepackage{graphicx}

\usepackage{amssymb}
\usepackage{amsmath}
\usepackage{paralist}
\usepackage{subfigure}

\usepackage{afterpage}
\usepackage{multirow}
\usepackage{multicol}
\usepackage{booktabs} 

\newcommand{\revision}[1]{#1}

\usepackage{amssymb}
\usepackage{amsfonts}
\usepackage{mathrsfs}
\usepackage{xspace}
\usepackage{bm}
\usepackage{upgreek}

\newcommand{\safemath}[2]{\newcommand{#1}{\ensuremath{#2}\xspace}}

\safemath{\bma}{\mathbf{a}}
\safemath{\bmb}{\mathbf{b}}
\safemath{\bmc}{\mathbf{c}}
\safemath{\bmd}{\mathbf{d}}
\safemath{\bme}{\mathbf{e}}
\safemath{\bmf}{\mathbf{f}}
\safemath{\bmg}{\mathbf{g}}
\safemath{\bmh}{\mathbf{h}}
\safemath{\bmi}{\mathbf{i}}
\safemath{\bmj}{\mathbf{j}}
\safemath{\bmk}{\mathbf{k}}
\safemath{\bml}{\mathbf{l}}
\safemath{\bmm}{\mathbf{m}}
\safemath{\bmn}{\mathbf{n}}
\safemath{\bmo}{\mathbf{o}}
\safemath{\bmp}{\mathbf{p}}
\safemath{\bmq}{\mathbf{q}}
\safemath{\bmr}{\mathbf{r}}
\safemath{\bms}{\mathbf{s}}
\safemath{\bmt}{\mathbf{t}}
\safemath{\bmu}{\mathbf{u}}
\safemath{\bmv}{\mathbf{v}}
\safemath{\bmw}{\mathbf{w}}
\safemath{\bmx}{\mathbf{x}}
\safemath{\bmy}{\mathbf{y}}
\safemath{\bmz}{\mathbf{z}}
\safemath{\bmzero}{\mathbf{0}}
\safemath{\bmone}{\mathbf{1}}

\bmdefine{\biad}{a}
\bmdefine{\bibd}{b}
\bmdefine{\bicd}{c}
\bmdefine{\bidd}{d}
\bmdefine{\bied}{e}
\bmdefine{\bifd}{f}
\bmdefine{\bigd}{g}
\bmdefine{\bihd}{h}
\bmdefine{\biid}{i}
\bmdefine{\bijd}{j}
\bmdefine{\bikd}{k}
\bmdefine{\bild}{l}
\bmdefine{\bimd}{m}
\bmdefine{\bind}{n}
\bmdefine{\biod}{o}
\bmdefine{\bipd}{p}
\bmdefine{\biqd}{q}
\bmdefine{\bird}{r}
\bmdefine{\bisd}{s}
\bmdefine{\bitd}{t}
\bmdefine{\biud}{u}
\bmdefine{\bivd}{v}
\bmdefine{\biwd}{w}
\bmdefine{\bixd}{x}
\bmdefine{\biyd}{y}
\bmdefine{\bizd}{z}

\bmdefine{\bixid}{\xi}
\bmdefine{\bilambdad}{\lambda}
\bmdefine{\bimud}{\mu}
\bmdefine{\bithetad}{\theta}
\bmdefine{\biphid}{\phi}
\bmdefine{\bideltad}{\delta}

\safemath{\bmia}{\biad}
\safemath{\bmib}{\bibd}
\safemath{\bmic}{\bicd}
\safemath{\bmid}{\bidd}
\safemath{\bmie}{\bied}
\safemath{\bmif}{\bifd}
\safemath{\bmig}{\bigd}
\safemath{\bmih}{\bihd}
\safemath{\bmii}{\biid}
\safemath{\bmij}{\bijd}
\safemath{\bmik}{\bikd}
\safemath{\bmil}{\bild}
\safemath{\bmim}{\bimd}
\safemath{\bmin}{\bind}
\safemath{\bmio}{\biod}
\safemath{\bmip}{\bipd}
\safemath{\bmiq}{\biqd}
\safemath{\bmir}{\bird}
\safemath{\bmis}{\bisd}
\safemath{\bmit}{\bitd}
\safemath{\bmiu}{\biud}
\safemath{\bmiv}{\bivd}
\safemath{\bmiw}{\biwd}
\safemath{\bmix}{\bixd}
\safemath{\bmiy}{\biyd}
\safemath{\bmiz}{\bizd}

\safemath{\bmxi}{\bixid}
\safemath{\bmlambda}{\bilambdad}
\safemath{\bmmu}{\bimud}
\safemath{\bmtheta}{\bithetad}
\safemath{\bmphi}{\biphid}
\safemath{\bmdelta}{\bideltad}

\safemath{\bA}{\mathbf{A}}
\safemath{\bB}{\mathbf{B}}
\safemath{\bC}{\mathbf{C}}
\safemath{\bD}{\mathbf{D}}
\safemath{\bE}{\mathbf{E}}
\safemath{\bF}{\mathbf{F}}
\safemath{\bG}{\mathbf{G}}
\safemath{\bH}{\mathbf{H}}
\safemath{\bI}{\mathbf{I}}
\safemath{\bJ}{\mathbf{J}}
\safemath{\bK}{\mathbf{K}}
\safemath{\bL}{\mathbf{L}}
\safemath{\bM}{\mathbf{M}}
\safemath{\bN}{\mathbf{N}}
\safemath{\bO}{\mathbf{O}}
\safemath{\bP}{\mathbf{P}}
\safemath{\bQ}{\mathbf{Q}}
\safemath{\bR}{\mathbf{R}}
\safemath{\bS}{\mathbf{S}}
\safemath{\bT}{\mathbf{T}}
\safemath{\bU}{\mathbf{U}}
\safemath{\bV}{\mathbf{V}}
\safemath{\bW}{\mathbf{W}}
\safemath{\bX}{\mathbf{X}}
\safemath{\bY}{\mathbf{Y}}
\safemath{\bZ}{\mathbf{Z}}

\safemath{\bZero}{\mathbf{0}}
\safemath{\bOne}{\mathbf{1}}
\safemath{\bDelta}{\mathbf{\Delta}}
\safemath{\bLambda}{\mathbf{\UpLambda}}
\safemath{\bPhi}{\mathbf{\Upphi}}
\safemath{\bSigma}{\mathbf{\Upsigma}}
\safemath{\bOmega}{\mathbf{\Upomega}}
\safemath{\bTheta}{\mathbf{\Uptheta}}

\bmdefine{\biAd}{A}
\bmdefine{\biBd}{B}
\bmdefine{\biCd}{C}
\bmdefine{\biDd}{D}
\bmdefine{\biEd}{E}
\bmdefine{\biFd}{F}
\bmdefine{\biGd}{G}
\bmdefine{\biHd}{H}
\bmdefine{\biId}{I}
\bmdefine{\biJd}{J}
\bmdefine{\biKd}{K}
\bmdefine{\biLd}{L}
\bmdefine{\biMd}{M}
\bmdefine{\biOd}{N}
\bmdefine{\biPd}{O}
\bmdefine{\biQd}{P}
\bmdefine{\biRd}{R}
\bmdefine{\biSd}{S}
\bmdefine{\biTd}{T}
\bmdefine{\biUd}{U}
\bmdefine{\biVd}{V}
\bmdefine{\biWd}{W}
\bmdefine{\biXd}{X}
\bmdefine{\biYd}{Y}
\bmdefine{\biZd}{Z}

\bmdefine{\biDelta}{\Delta}
\bmdefine{\biLambda}{\Lambda}
\bmdefine{\biPhi}{\Phi}
\bmdefine{\biSigma}{\Sigma}
\bmdefine{\biOmega}{\Omega}
\bmdefine{\biTheta}{\Theta}

\safemath{\bimA}{\biAd}
\safemath{\bimB}{\biBd}
\safemath{\bimC}{\biCd}
\safemath{\bimD}{\biDd}
\safemath{\bimE}{\biEd}
\safemath{\bimF}{\biFd}
\safemath{\bimG}{\biGd}
\safemath{\bimH}{\biHd}
\safemath{\bimI}{\biId}
\safemath{\bimJ}{\biJd}
\safemath{\bimK}{\biKd}
\safemath{\bimL}{\biLd}
\safemath{\bimM}{\biMd}
\safemath{\bimN}{\biNd}
\safemath{\bimO}{\biOd}
\safemath{\bimP}{\biPd}
\safemath{\bimQ}{\biQd}
\safemath{\bimR}{\biRd}
\safemath{\bimS}{\biSd}
\safemath{\bimT}{\biTd}
\safemath{\bimU}{\biUd}
\safemath{\bimV}{\biVd}
\safemath{\bimW}{\biWd}
\safemath{\bimX}{\biXd}
\safemath{\bimY}{\biYd}
\safemath{\bimZ}{\biZd}

\safemath{\bimDelta}{\biDelta}
\safemath{\bimLambda}{\biLambda}
\safemath{\bimPhi}{\biPhi}
\safemath{\bimSigma}{\biSigma}
\safemath{\bimOmega}{\biOmega}
\safemath{\bimTheta}{\biTheta}

\safemath{\setA}{\mathcal{A}}
\safemath{\setB}{\mathcal{B}}
\safemath{\setC}{\mathcal{C}}
\safemath{\setD}{\mathcal{D}}
\safemath{\setE}{\mathcal{E}}
\safemath{\setF}{\mathcal{F}}
\safemath{\setG}{\mathcal{G}}
\safemath{\setH}{\mathcal{H}}
\safemath{\setI}{\mathcal{I}}
\safemath{\setJ}{\mathcal{J}}
\safemath{\setK}{\mathcal{K}}
\safemath{\setL}{\mathcal{L}}
\safemath{\setM}{\mathcal{M}}
\safemath{\setN}{\mathcal{N}}
\safemath{\setO}{\mathcal{O}}
\safemath{\setP}{\mathcal{P}}
\safemath{\setQ}{\mathcal{Q}}
\safemath{\setR}{\mathcal{R}}
\safemath{\setS}{\mathcal{S}}
\safemath{\setT}{\mathcal{T}}
\safemath{\setU}{\mathcal{U}}
\safemath{\setV}{\mathcal{V}}
\safemath{\setW}{\mathcal{W}}
\safemath{\setX}{\mathcal{X}}
\safemath{\setY}{\mathcal{Y}}
\safemath{\setZ}{\mathcal{Z}}
\safemath{\emptySet}{\varnothing}

\safemath{\colA}{\mathscr{A}}
\safemath{\colB}{\mathscr{B}}
\safemath{\colC}{\mathscr{C}}
\safemath{\colD}{\mathscr{D}}
\safemath{\colE}{\mathscr{E}}
\safemath{\colF}{\mathscr{F}}
\safemath{\colG}{\mathscr{G}}
\safemath{\colH}{\mathscr{H}}
\safemath{\colI}{\mathscr{I}}
\safemath{\colJ}{\mathscr{J}}
\safemath{\colK}{\mathscr{K}}
\safemath{\colL}{\mathscr{L}}
\safemath{\colM}{\mathscr{M}}
\safemath{\colN}{\mathscr{N}}
\safemath{\colO}{\mathscr{O}}
\safemath{\colP}{\mathscr{P}}
\safemath{\colQ}{\mathscr{Q}}
\safemath{\colR}{\mathscr{R}}
\safemath{\colS}{\mathscr{S}}
\safemath{\colT}{\mathscr{T}}
\safemath{\colU}{\mathscr{U}}
\safemath{\colV}{\mathscr{V}}
\safemath{\colW}{\mathscr{W}}
\safemath{\colX}{\mathscr{X}}
\safemath{\colY}{\mathscr{Y}}
\safemath{\colZ}{\mathscr{Z}}

\safemath{\opA}{\mathbb{A}}
\safemath{\opB}{\mathbb{B}}
\safemath{\opC}{\mathbb{C}}
\safemath{\opD}{\mathbb{D}}
\safemath{\opE}{\mathbb{E}}
\safemath{\opF}{\mathbb{F}}
\safemath{\opG}{\mathbb{G}}
\safemath{\opH}{\mathbb{H}}
\safemath{\opI}{\mathbb{I}}
\safemath{\opJ}{\mathbb{J}}
\safemath{\opK}{\mathbb{K}}
\safemath{\opL}{\mathbb{L}}
\safemath{\opM}{\mathbb{M}}
\safemath{\opN}{\mathbb{N}}
\safemath{\opO}{\mathbb{O}}
\safemath{\opP}{\mathbb{P}}
\safemath{\opQ}{\mathbb{Q}}
\safemath{\opR}{\mathbb{R}}
\safemath{\opS}{\mathbb{S}}
\safemath{\opT}{\mathbb{T}}
\safemath{\opU}{\mathbb{U}}
\safemath{\opV}{\mathbb{V}}
\safemath{\opW}{\mathbb{W}}
\safemath{\opX}{\mathbb{X}}
\safemath{\opY}{\mathbb{Y}}
\safemath{\opZ}{\mathbb{Z}}
\safemath{\opZero}{\mathbb{O}}
\safemath{\identityop}{\opI}

\safemath{\veca}{\bma}
\safemath{\vecb}{\bmb}
\safemath{\vecc}{\bmc}
\safemath{\vecd}{\bmd}
\safemath{\vece}{\bme}
\safemath{\vecf}{\bmf}
\safemath{\vecg}{\bmg}
\safemath{\vech}{\bmh}
\safemath{\veci}{\bmi}
\safemath{\vecj}{\bmj}
\safemath{\veck}{\bmk}
\safemath{\vecl}{\bml}
\safemath{\vecm}{\bmm}
\safemath{\vecn}{\bmn}
\safemath{\veco}{\bmo}
\safemath{\vecp}{\bmp}
\safemath{\vecq}{\bmq}
\safemath{\vecr}{\bmr}
\safemath{\vecs}{\bms}
\safemath{\vect}{\bmt}
\safemath{\vecu}{\bmu}
\safemath{\vecv}{\bmv}
\safemath{\vecw}{\bmw}
\safemath{\vecx}{\bmx}
\safemath{\vecy}{\bmy}
\safemath{\vecz}{\bmz}

\safemath{\veczero}{\bmzero}
\safemath{\vecone}{\bmone}
\safemath{\vecxi}{\bmxi}
\safemath{\veclambda}{\bmlambda}
\safemath{\vecmu}{\bmmu}
\safemath{\vectheta}{\bmtheta}
\safemath{\vecphi}{\bmphi}
\safemath{\vecdelta}{\bmdelta}

\safemath{\matA}{\bA}
\safemath{\matB}{\bB}
\safemath{\matC}{\bC}
\safemath{\matD}{\bD}
\safemath{\matE}{\bE}
\safemath{\matF}{\bF}
\safemath{\matG}{\bG}
\safemath{\matH}{\bH}
\safemath{\matI}{\bI}
\safemath{\matJ}{\bJ}
\safemath{\matK}{\bK}
\safemath{\matL}{\bL}
\safemath{\matM}{\bM}
\safemath{\matN}{\bN}
\safemath{\matO}{\bO}
\safemath{\matP}{\bP}
\safemath{\matQ}{\bQ}
\safemath{\matR}{\bR}
\safemath{\matS}{\bS}
\safemath{\matT}{\bT}
\safemath{\matU}{\bU}
\safemath{\matV}{\bV}
\safemath{\matW}{\bW}
\safemath{\matX}{\bX}
\safemath{\matY}{\bY}
\safemath{\matZ}{\bZ}
\safemath{\matzero}{\bmzero}

\safemath{\matDelta}{\bDelta}
\safemath{\matLambda}{\bLambda}
\safemath{\matPhi}{\bPhi}
\safemath{\matSigma}{\bSigma}
\safemath{\matOmega}{\bOmega}
\safemath{\matTheta}{\bTheta}

\safemath{\matidentity}{\matI}
\safemath{\matone}{\matO}

\safemath{\rnda}{A}
\safemath{\rndb}{B}
\safemath{\rndc}{C}
\safemath{\rndd}{D}
\safemath{\rnde}{E}
\safemath{\rndf}{F}
\safemath{\rndg}{G}
\safemath{\rndh}{H}
\safemath{\rndi}{I}
\safemath{\rndj}{J}
\safemath{\rndk}{K}
\safemath{\rndl}{L}
\safemath{\rndm}{M}
\safemath{\rndn}{N}
\safemath{\rndo}{O}
\safemath{\rndp}{P}
\safemath{\rndq}{Q}
\safemath{\rndr}{R}
\safemath{\rnds}{S}
\safemath{\rndt}{T}
\safemath{\rndu}{U}
\safemath{\rndv}{V}
\safemath{\rndw}{W}
\safemath{\rndx}{X}
\safemath{\rndy}{Y}
\safemath{\rndz}{Z}

\safemath{\rveca}{\bimA}
\safemath{\rvecb}{\bimB}
\safemath{\rvecc}{\bimC}
\safemath{\rvecd}{\bimD}
\safemath{\rvece}{\bimE}
\safemath{\rvecf}{\bimF}
\safemath{\rvecg}{\bimG}
\safemath{\rvech}{\bimH}
\safemath{\rveci}{\bimI}
\safemath{\rvecj}{\bimJ}
\safemath{\rveck}{\bimK}
\safemath{\rvecl}{\bimL}
\safemath{\rvecm}{\bimM}
\safemath{\rvecn}{\bimN}
\safemath{\rveco}{\bomO}
\safemath{\rvecp}{\bimP}
\safemath{\rvecq}{\bimQ}
\safemath{\rvecr}{\bimR}
\safemath{\rvecs}{\bimS}
\safemath{\rvect}{\bimT}
\safemath{\rvecu}{\bimU}
\safemath{\rvecv}{\bimV}
\safemath{\rvecw}{\bimW}
\safemath{\rvecx}{\bimX}
\safemath{\rvecy}{\bimY}
\safemath{\rvecz}{\bimZ}

\safemath{\rvecxi}{\bmxi}
\safemath{\rveclambda}{\bmlambda}
\safemath{\rvecmu}{\bmmu}
\safemath{\rvectheta}{\bmtheta}
\safemath{\rvecphi}{\bmphi}

\safemath{\rmatA}{\bimA}
\safemath{\rmatB}{\bimB}
\safemath{\rmatC}{\bimC}
\safemath{\rmatD}{\bimD}
\safemath{\rmatE}{\bimE}
\safemath{\rmatF}{\bimF}
\safemath{\rmatG}{\bimG}
\safemath{\rmatH}{\bimH}
\safemath{\rmatI}{\bimI}
\safemath{\rmatJ}{\bimJ}
\safemath{\rmatK}{\bimK}
\safemath{\rmatL}{\bimL}
\safemath{\rmatM}{\bimM}
\safemath{\rmatN}{\bimN}
\safemath{\rmatO}{\bimO}
\safemath{\rmatP}{\bimP}
\safemath{\rmatQ}{\bimQ}
\safemath{\rmatR}{\bimR}
\safemath{\rmatS}{\bimS}
\safemath{\rmatT}{\bimT}
\safemath{\rmatU}{\bimU}
\safemath{\rmatV}{\bimV}
\safemath{\rmatW}{\bimW}
\safemath{\rmatX}{\bimX}
\safemath{\rmatY}{\bimY}
\safemath{\rmatZ}{\bimZ}

\safemath{\rmatDelta}{\bimDelta}
\safemath{\rmatLambda}{\bimLambda}
\safemath{\rmatPhi}{\bimPhi}
\safemath{\rmatSigma}{\bimSigma}
\safemath{\rmatOmega}{\bimOmega}
\safemath{\rmatTheta}{\bimTheta}

\usepackage{amssymb}
\usepackage{amsfonts}
\usepackage{mathrsfs}
\usepackage{xspace}
\usepackage{bm}
\usepackage{fancyref}
\usepackage{textcomp}

\usepackage{multirow}
\usepackage{stmaryrd}

\newenvironment{textbmatrix}{	\setlength{\arraycolsep}{2.5pt}%
								\big[\begin{matrix}}{\end{matrix}\big]%
								\raisebox{0.08ex}{\vphantom{M}}}

\def\be{\begin{equation}}
\def\ee{\end{equation}}
\def\een{\nonumber \end{equation}}
\def\mat{\begin{bmatrix}}
\def\emat{\end{bmatrix}}
\def\btm{\begin{textbmatrix}}
\def\etm{\end{textbmatrix}}

\def\ba#1\ea{\begin{align}#1\end{align}}
\def\bas#1\eas{\begin{align*}#1\end{align*}}
\def\bs#1\es{\begin{split}#1\end{split}} 
\def\bg#1\eg{\begin{gather}#1\end{gather}}
\def\bml#1\eml{\begin{multline}#1\end{multline}}
\def\bi#1\ei{\begin{itemize}#1\end{itemize}}

\newcommand{\lefto}{\mathopen{}\left}

\DeclareMathOperator*{\argmin}{arg\;min}		

\newcommand{\abs}[1]{\lefto\lvert#1\right\rvert}		

\newcommand{\vecnorm}[1]{\lefto\lVert#1\right\rVert}		
\newcommand{\herm}[1]{\ensuremath{#1^{H}}} 	

\safemath{\dirac}{\delta}					
\safemath{\krond}{\dirac}					

\safemath{\upto}{\uparrow}
\safemath{\downto}{\downarrow}
\safemath{\iu}{j}							
\safemath{\ev}{\lambda}						
\safemath{\hilseqspace}{l^{2}}				
\newcommand{\banachfunspace}[1]{\setL^{#1}}	
\safemath{\hilfunspace}{\banachfunspace{2}}	

\safemath{\SNR}{\text{\sc snr}} 				
\safemath{\No}{N_0}							
\safemath{\Es}{E_s}							
\safemath{\Eb}{E_b}							
\safemath{\EbNo}{\frac{\Eb}{\No}}
\safemath{\EsNo}{\frac{\Es}{\No}}

\DeclareMathOperator{\CHop}{\ensuremath{\opH}} 
\safemath{\tvir}{\rndh_{\CHop}}				
\safemath{\tvtf}{\rndl_{\CHop}}				
\safemath{\spf}{\rnds_{\CHop}}				
\safemath{\bff}{H_{\CHop}}					

\safemath{\ircf}{r_{h}}						
\safemath{\tftvcf}{r_{s}}					
\safemath{\tfcf}{r_{l}}						
\safemath{\bfcf}{r_{H}}						

\safemath{\tcorr}{c_h}						
\safemath{\scf}{c_{s}}						
\safemath{\tfcorr}{c_{l}}					
\safemath{\fcorr}{c_{H}}						

\safemath{\mi}{I}							
\safemath{\capacity}{C}						

\safemath{\normal}{\mathcal{N}}			
\safemath{\jpg}{\mathcal{CN}}			
\safemath{\mchain}{\leftrightarrow}		

\safemath{\dB}{\,\mathrm{dB}}
\safemath{\dBm}{\,\mathrm{dBm}}
\safemath{\Hz}{\,\mathrm{Hz}}
\safemath{\kHz}{\,\mathrm{kHz}}
\safemath{\MHz}{\,\mathrm{MHz}}
\safemath{\GHz}{\,\mathrm{GHz}}
\safemath{\s}{\,\mathrm{s}}
\safemath{\ms}{\,\mathrm{ms}}
\safemath{\mus}{\,\mathrm{\text{\textmu}s}}
\safemath{\ns}{\,\mathrm{ns}}
\safemath{\ps}{\,\mathrm{ps}}
\safemath{\meter}{\,\mathrm{m}}
\safemath{\mm}{\,\mathrm{mm}}
\safemath{\cm}{\,\mathrm{cm}}
\safemath{\m}{\,\mathrm{m}}
\safemath{\W}{\,\mathrm{W}}
\safemath{\mW}{\, \mathrm{mW}}
\safemath{\J}{\,\mathrm{J}}
\safemath{\K}{\,\mathrm{K}}
\safemath{\bit}{\,\mathrm{bit}}
\safemath{\nat}{\,\mathrm{nat}}

\safemath{\define}{\triangleq}			

\safemath{\equivalent}{\sim}
\safemath{\distas}{\sim}					
\safemath{\sdiff}{\Delta}				

\safemath{\reals}{\mathbb{R}}
\safemath{\positivereals}{\reals_{+}}
\safemath{\integers}{\mathbb{Z}}
\safemath{\posint}{\integers_{+}}
\safemath{\naturals}{\mathbb{N}}
\safemath{\posnaturals}{\naturals_{+}}
\safemath{\complexset}{\mathbb{C}}
\safemath{\rationals}{\mathbb{Q}}

\newcommand*{\fancyrefapplabelprefix}{app}		
\newcommand*{\fancyrefthmlabelprefix}{thm}		
\newcommand*{\fancyreflemlabelprefix}{lem}		
\newcommand*{\fancyrefcorlabelprefix}{cor}		
\newcommand*{\fancyrefdeflabelprefix}{def}		
\newcommand*{\fancyrefproplabelprefix}{prop}		
\newcommand*{\fancyrefexmpllabelprefix}{exmpl}
\frefformat{vario}{\fancyrefseclabelprefix}{Section~#1}
\frefformat{vario}{\fancyrefthmlabelprefix}{Theorem~#1}
\frefformat{vario}{\fancyreflemlabelprefix}{Lemma~#1}
\frefformat{vario}{\fancyrefcorlabelprefix}{Corollary~#1}
\frefformat{vario}{\fancyrefdeflabelprefix}{Definition~#1}
\frefformat{vario}{\fancyreffiglabelprefix}{Fig.~#1}
\frefformat{vario}{\fancyrefapplabelprefix}{#1} 
\frefformat{vario}{\fancyrefeqlabelprefix}{(#1)}
\frefformat{vario}{\fancyrefproplabelprefix}{Property~#1}
\frefformat{vario}{\fancyrefexmpllabelprefix}{Example~#1}
\frefformat{vario}{\fancyreftablabelprefix}{Table~#1}

 \newtheorem{thm}{Theorem}
 
 \newtheorem{defi}{Definition}
 \newtheorem{lem}[thm]{Lemma}
 
\newcommand{\RIC}[1]{\delta_{#1}}
\safemath{\BPDN}{\textrm{BPDN}}

\safemath{\ysig}{\bmy}
\safemath{\ysighat}{\hat{\ysig}}
\safemath{\ysigdim}{M}
\safemath{\xsig}{\bmx}
\safemath{\xsigdim}{N}
\safemath{\nx}{n_x}
\safemath{\zsig}{\bmz}
\safemath{\zsigdim}{\ysigdim}
\safemath{\rsig}{\bmr}
\safemath{\Adict}{\bA}
\safemath{\Adicttilde}{\widetilde{\Adict}}
\safemath{\Adictdim}{\outputdim\times\xsigdim}
\safemath{\avec}{\bma}
\safemath{\avectilde}{\tilde{\avec}}
\safemath{\Bdict}{\bB}
\safemath{\Bdicttilde}{\widetilde{\Bdict}}
\safemath{\Cdict}{\bC}
\safemath{\cvec}{\bmc}
\safemath{\Ddict}{\bD}
\safemath{\Ddictdim}{\ysigdim\times\xsigdim}
\safemath{\dvec}{\bmd}
\safemath{\Ddicttilde}{\widetilde{\bD}}
\safemath{\Bonb}{\bB}
\safemath{\bvec}{\bmb}
\safemath{\Bonbdim}{\ysigdim\times\ysigdim}
\safemath{\noise}{\bmn}
\safemath{\noisedim}{\ysigim}
\safemath{\err}{\bme}
\safemath{\errdim}{\ysigdim}
\safemath{\errset}{\setE}
\safemath{\nerr}{n_e}
\safemath{\delop}{\bP_\errset}
\safemath{\delopc}{\bP_{{\errset}^c}}

\safemath{\cplxi}{\imath}
\safemath{\cplxj}{\jmath}

\safemath{\dict}{\matD}
\safemath{\inputdim}{N}		
\safemath{\outputdim}{M}		
\safemath{\sparsity}{S}	
\safemath{\inputdimA}{{N_a}}	
\safemath{\inputdimB}{{N_b}}	
\safemath{\elemA}{{n_a}}	
\safemath{\elemB}{{n_b}}	
\safemath{\resA}{\matR_a}	
\safemath{\resB}{\matR_b}	
\safemath{\subD}{\matS} 
\safemath{\subA}{\matS_a} 
\safemath{\subB}{\matS_b} 
\safemath{\dicta}{\matA} 	
\safemath{\dictb}{\matB} 	
\safemath{\hollowS}{H}
\safemath{\hollowA}{H_a}
\safemath{\hollowB}{H_b}
\safemath{\cross}{Z}
\safemath{\coh}{\mu_d}			
\safemath{\coha}{\mu_a}			
\safemath{\cohb}{\mu_b}			
\safemath{\mubs}{\nu}	
\safemath{\cohm}{\mu_m} 
\safemath{\dictset}{\setD}	
\safemath{\dictsetp}{\dictset(\coh,\coha,\cohb)}	
\safemath{\dictsetgen}{\dictset_\text{gen}}
\safemath{\dictsetgenp}{\dictsetgen(\coh)}
\safemath{\dictsetonb}{\dictset_\text{onb}}
\safemath{\dictsetonbp}{\dictsetonb(\coh)}

\safemath{\leftside}{U}
\safemath{\rightsideA}{R_a}
\safemath{\rightsideB}{R_b}

\safemath{\indexS}{\setI_S} 

\safemath{\na}{n_a}			
\safemath{\nb}{n_b}			
\safemath{\coeffa}{p_i}	
\safemath{\coeffb}{q_j}	
\safemath{\seta}{\setP}		
\safemath{\setb}{\setQ}     
\safemath{\setw}{\setW}	
\safemath{\setz}{\setZ}	
\safemath{\cola}{\veca}		
\safemath{\colb}{\vecb}		
\safemath{\cold}{\vecd}		
\safemath{\inputvec}{\vecx} 	
\safemath{\error}{\vece}	
\safemath{\noiseout}{\vecz} 	
\safemath{\inputvecel}{x}
\safemath{\inputveca}{\vecx_a}
\safemath{\inputvecb}{\vecx_b}
\safemath{\outputvec}{\vecy}	
\safemath{\lambdamin}{\lambda_{\mathrm{min}}}

\newcommand{\pos}[1]{\lefto[#1\right]^+}
\newcommand{\normtwo}[1]{\vecnorm{#1}_2}
\newcommand{\normone}[1]{\vecnorm{#1}_1}
\newcommand{\normzero}[1]{\vecnorm{#1}_0}

\newcommand{\normfro}[1]{\vecnorm{#1}_F}

\safemath{\elltwo}{\ell_2}
\safemath{\ellone}{\ell_1}
\safemath{\ellzero}{\ell_0}
\safemath{\ellinf}{\ell_\infty}
\safemath{\licard}{Z(\coh,\coha,\cohb)}
\safemath{\xsol}{\hat{x}}
\safemath{\xbord}{x_b}		
\safemath{\xstat}{x_s}		
\safemath{\xstatLone}{\tilde{x}_s}
\safemath{\order}{\mathcal{O}} 
\safemath{\scales}{\Theta} 
\safemath{\ones}{\mathbf{1}} 
\safemath{\zeroes}{\mathbf{0}} 
\safemath{\thlone}{\kappa(\coh,\cohb)} 
\safemath{\constoneA}{\delta} 
\safemath{\constoneB}{\epsilon} 
\safemath{\nlarge}{L}				   
\safemath{\sumlarge}{S_\nlarge}
\safemath{\maxlarger}{P_\nlarge}	   
\safemath{\Pzero}{\textrm{P0}}	
\safemath{\Pone}{\textrm{P1}}

\safemath{\vecfir}{\vecw}			 
\safemath{\vecsec}{\vecz}
\safemath{\elvecfir}{w}              
\safemath{\elvecsec}{z}				 
\safemath{\nlargefir}{n}
\safemath{\normout}{\gamma}
\safemath{\auxfun}{h}
\safemath{\supp}{\textrm{supp}}

\safemath{\indexa}{\ell}
\safemath{\indexb}{r}
\safemath{\indexc}{i}
\safemath{\indexd}{j}

\safemath{\project}{P}

\newproof{IEEEproof}{Proof}

\journal{Applied and Computational Harmonic Analysis}

\begin{document}

\begin{frontmatter}

\title{{\bf Stable Restoration and Separation of Approximately Sparse Signals}}


\author{Christoph Studer\corref{cor1}\fnref{fn1}} \ead{studer@rice.edu} \ead[url]{http://www.ece.rice.edu/~cs32/}
\author{Richard G.\ Baraniuk\corref{cor2}\fnref{fn2}} \ead{richb@rice.edu} \ead[url]{http://web.ece.rice.edu/richb/}

\address{Dept.~of Electrical and Computer Engineering, Rice University, MS-380 \\ 6100 Main Street, Houston, TX 77005, USA \\ }

\cortext[cor1]{Corresponding author}
\cortext[cor2]{Principal corresponding author}

\fntext[fn1]{Phone: +1 713.348.3579; Fax: +1 713.348.5685}
\fntext[fn2]{Phone: +1 713.348.5132; Fax: +1 713.348.5685}



\begin{abstract}
This paper develops new theory and algorithms to recover signals that are approximately sparse in some general dictionary (i.e., a basis, frame, or over-/incomplete matrices) but corrupted by a combination of interference having a sparse representation in a second general dictionary and measurement noise.  
\revision{The algorithms and analytical recovery conditions consider varying degrees of signal and interference support-set knowledge.}
Particular applications covered by the proposed framework include the restoration of signals impaired by  impulse noise, narrowband interference, or saturation/clipping, as well as image in-painting, super-resolution, and signal separation.
Two application examples for audio and image restoration demonstrate the efficacy of the approach. 
\end{abstract}


\begin{keyword}
Sparse signal recovery \sep signal restoration \sep signal separation \sep deterministic recovery guarantees \sep coherence \sep basis-pursuit denoising
\end{keyword}

\end{frontmatter}


\section{Introduction}


We investigate the \emph{recovery}  problem of the coefficient vector~$\inputvec\in\complexset^\inputdimA$ from the corrupted~$\outputdim$-dimensional observations
\begin{align} \label{eq:systemmodel}
 \noiseout = \dicta \inputvec + \dictb\error + \noise,
\end{align}
where $\dicta\in\complexset^{\outputdim\times \inputdimA}$ and $\dictb\in\complexset^{\outputdim\times \inputdimB}$ are general deterministic dictionaries; examples for general dictionaries include bases, frames, or over-/incomplete matrices whose columns have unit Euclidean (or \elltwo) norm. 
The vector \inputvec is assumed to be \emph{approximately sparse}, i.e., its main energy (in terms of the sum of absolute values, \revision{for example}) is concentrated in only a few entries. The $M$-dimensional signal vector is defined as $\outputvec=\dicta\inputvec$.
The vector $\error\in\complexset^\inputdimA$ represents interference and is assumed to be \emph{perfectly sparse}, i.e., only a few entries are nonzero, and $\noise\in\complexset^\outputdim$ corresponds to measurement noise. Apart from the bound $\normtwo{\noise}<\varepsilon$, the measurement noise is arbitrary.
We emphasize that the interference and noise components \error and \noise can depend on the vector~\inputvec and/or the dictionary~\dicta.

The setting \fref{eq:systemmodel} also allows us to study \emph{signal separation}, i.e., the separation of two distinct features $\dicta\inputvec$ and $\dictb\error$  from the noisy observation~\noiseout. Here, the vector \error in \fref{eq:systemmodel} is also allowed to be approximately sparse and is used to represent a second desirable feature (rather than undesired interference).
Signal separation amounts to simultaneously recovering the vectors \inputvec and~\error from the noisy measurement~\noiseout followed by computation of the individual signal features $\dicta\inputvec$ and $\dictb\error$.

\subsection{Applications for the model \fref{eq:systemmodel}}

Both the recovery and separation problems outlined above feature prominently in numerous applications (see \cite{VF92,GR98,elad2001,elad2005,GH07,cai2009,laska2009democracy,bertalmio2000,WO08,CCSS09,laska,SKPB10,carrillo2010,novak2010,mallat2010,Adler2011,AEJEGP11,Kutyniok:2011fk} and the references therein), including:
\begin{itemize}

\item \emph{Impulse noise:} The recovery of approximately sparse signals corrupted by impulse noise~\cite{carrillo2010}
corresponds to recovery of \inputvec from~\fref{eq:systemmodel} by setting $\dictb=\bI_\outputdim$ and associating the interference \error with the impulse-noise vector.
Practical examples include restoration of audio signals impaired by click/pop noise~\cite{VF92,GR98} and reading from unreliable memories~\cite{novak2010}.

\item \emph{Narrowband interference:} Audio, video, and communication signals are often corrupted by narrowband interference. A particular example is electric hum, which typically occurs in improperly designed audio or video equipment.
Such impairments naturally exhibit a sparse representation in the frequency domain, which amounts to setting $\dictb$ to the inverse discrete Fourier transform matrix. 

\item \emph{Saturation  and clipping:} Non-linearities in amplifiers may result in signal saturation, cf.~\cite{laska2009democracy,Adler2011,AEJEGP11}.
Here, instead of the signal vector \outputvec of interest, one observes a saturated (or clipped) version $\noiseout=\outputvec + \error + \noise$, where the nonzero entries of $\error$ correspond to the difference between the saturated signal and the original signal $\outputvec$. The noise vector \noise can be used to model residual errors that are not captured by the interference component~$\dictb\error$.

\item \emph{Super-resolution and in-painting:} 
In super-resolution~\cite{mallat2010,elad2001} and in-painting~\cite{bertalmio2000,WO08,CCSS09,cai2009} applications, only a subset of the entries of the (full-resolution) signal vector $\outputvec=\dicta\inputvec$ is available.
With \fref{eq:systemmodel}, the interference vector \error accounts for the missing parts of the signal, i.e., the locations of the nonzero entries of \error correspond to the missing entries in~$\outputvec$ and are set to some arbitrary value.
The missing parts of~\outputvec are then filled in by recovering \inputvec from $\noiseout=\dicta\inputvec+\error + \noise$ followed by computation of the (full-resolution) signal vector $\outputvec=\dicta\inputvec$. 

\item \emph{Signal separation:} 
The framework~\fref{eq:systemmodel} can be used to model the decomposition of signals into two distinct features. 
Prominent application examples are the separation of texture from cartoon parts in images~\cite{elad2005,cai2009,Kutyniok:2011fk} and the separation of neuronal calcium transients from smooth signals caused by astrocytes in calcium imaging~\cite{GH07}.
In both applications, \dicta and \dictb  are chosen such that each feature can be represented by approximately sparse vectors in one dictionary.
Signal separation then amounts to simultaneously extracting \inputvec and~\error from $\noiseout$, where $\dicta\inputvec$ and $\dictb\error$ represent the individual features.

\end{itemize}

In many applications outlined above, a predetermined (and possibly optimized) dictionary pair \dicta and \dictb is used. 
It is therefore of significant practical interest to identify the fundamental limits on the performance of restoration or separation from the model \fref{eq:systemmodel} for the deterministic setting, i.e., assuming no randomness in the dictionaries, the signal, interference, or the noise vector. 
Deterministic recovery guarantees for the special case of {\em perfectly} sparse vectors \inputvec and \error and {\em no} measurement noise have been studied in~\cite{SKPB10,kuppinger2010a}. 
The results in \cite{SKPB10,kuppinger2010a} rely on an uncertainty relation for pairs of general dictionaries and depend on the number of nonzero entries in \inputvec and \error, on the coherence parameters of the dictionaries \dicta and \dictb, and on the amount of prior knowledge on the support of the signal and interference vector.
However, the algorithms and proof techniques used in~\cite{SKPB10,kuppinger2010a} cannot be adapted for the general (and practically more relevant) setting formulated in~\fref{eq:systemmodel}, which features approximately sparse signals and additive measurement noise.

\subsection{Contributions}
In this paper, we generalize the recovery guarantees of~\cite{SKPB10,kuppinger2010a} to the framework~\fref{eq:systemmodel} detailed above.
\revision{In particular, we provide computationally efficient restoration and separation algorithms and derive corresponding recovery guarantees for the deterministic setting.}
Our guarantees depend in a natural way on the number of dominant nonzero entries of \inputvec and \error, on the coherence parameters of the dictionaries \dicta and \dictb, and on the Euclidean norm of the measurement noise.
Our results also depend on the amount of  knowledge on the location of the dominant entries available prior to recovery. In particular, we investigate the following cases: 
\begin{inparaenum}[1)]
\item The locations of the dominant entries of the approximately sparse vector \inputvec and the support set of the perfectly sparse interference vector \error are known (prior to recovery),
\item only the support set of the interference vector \error is known, and
\item no support-set knowledge about \inputvec and \error is available.
\end{inparaenum}
\revision{Moreover, we present coherence-based bounds on the restricted isometry constants (RICs) for all these cases, which can be used to derive alternative recovery conditions.}
We provide a comparison to the recovery conditions for perfectly sparse signals and noiseless measurements presented in~\cite{SKPB10,kuppinger2010a}.
Finally, we demonstrate the efficacy of the proposed approach with two representative applications: restoration of audio signals impaired by a mixture of impulse noise and Gaussian noise, and removal of scratches from \revision{color photographs.}

\subsection{Notation}
Lowercase and uppercase boldface letters stand for column vectors and matrices, respectively. 
The transpose, conjugate transpose, and (Moore--Penrose) pseudo-inverse of the matrix \bM are denoted by $\bM^T$, $\herm{\bM}$, and $\bM^{\dagger}=\left(\bM^H\bM\right)^{\!-1}\bM^H$, respectively. 
The $k$th entry of the vector $\bmm$ is $[\vecm]_k$, and the $k$th column of \bM is $\bmm_k$ and the entry in the $k$th row and $\ell$th column is designated by $[\bM]_{k,\ell}$. 
The $M\times M$ identity matrix is denoted by $\bI_M$ and the $M\times N$ all zeros matrix by $\mathbf{0}_{M\times{}N}$. 
The Euclidean (or \elltwo) norm of the vector $\bmx$ is denoted by $\normtwo{\bmx}$, \mbox{$\normone{\bmx} = \sum_{k} \abs{[\bmx]_k}$} stands for the \ellone-norm of \inputvec, and $\normzero{\bmx}$ designates the number of nonzero entries of \bmx. 
The spectral norm of the matrix \bM is $\normtwo{\bM}=\sqrt{\lambda_\text{max}(\bM^H\bM)}$, where the minimum and maximum eigenvalue of a positive-semidefinite matrix \bM are denoted by \mbox{$\lambda_{\textrm{min}}(\bM)$} and $\lambda_\textrm{max}(\bM)$, respectively. $\normfro{\bM}=\sqrt{\sum_{k,\ell}\abs{[\bM]_{k,\ell}}^2}$ stands for the Frobenius matrix norm.
Sets are designated by upper-case calligraphic letters. The cardinality of the set \setT is $\abs{\setT}$ and the complement of a set~$\setS$ in some superset~$\setT$ is denoted by $\setS^c$.
The support set of the vector \vecm, i.e., the index set corresponding to the nonzero entries of \vecm, is designated by $\supp(\vecm)$. 
We define the $M\times M$ diagonal (projection) matrix~$\bP_\setS$ for the set $\setS\subseteq\{1,\ldots,M\}$ as follows:
\begin{align*}
[\bP_\setS]_{k,\ell} = 
\left\{\begin{array}{ll}
1, & k=\ell \text{ and } k \in \setS \\
0, & \textrm{otherwise,}
\end{array}\right.
\end{align*}
and $\vecm_\setT = \bP_\setT\vecm$.
The matrix $\bM_\setT$ is obtained from \bM by retaining the columns of \bM with indices in \setT and the $\abs{\setT}$-dimensional vector $[\bmm]_\setT$ is obtained analogously.
For $x\in\reals$,  we set $\pos{x}\!=\max\{x,0\}$.

\subsection{Synopsis}
The remainder of the paper is organized as follows. In \fref{sec:priorart}, we briefly summarize the relevant prior art. Our new recovery algorithms and corresponding recovery guarantees are presented in \fref{sec:mainresults}. \revision{A set of alternative recovery guarantees obtained through the restricted isometry property (RIP) framework and a comparison to existing recovery guarantees are provided in \fref{sec:discussion}.} The application examples are shown in \fref{sec:application}, and we conclude in \fref{sec:conclusions}. \revision{All proofs are relegated to the appendices.}


\section{Relevant Prior Art}
\label{sec:priorart}

In this section, we review the relevant prior art in recovering sparse signals from noiseless and noisy measurements in the deterministic setting and summarize the existing guarantees for recovery of sparsely corrupted signals. 

\subsection{Recovery of perfectly sparse signals from noiseless measurements} 
\label{sec:noiselesscase}

Recovery of a vector $\inputvec\in\complexset^\inputdimA$ from the noiseless observations $\outputvec=\dicta\inputvec$ with~\dicta over-complete (i.e., $\outputdim<N_a$) corresponds to solving an underdetermined system of linear equations, which is well-known to be ill-posed. 
However, assuming that \inputvec is perfectly sparse (i.e., that only small number of its entries are nonzero) enables one to uniquely recover~\inputvec by solving
\begin{align*}
(\text{P0}) & \quad \mathit{minimize\;} 
\normzero{\tilde\inputvec} \quad \mathit{subject\;to} \;  \outputvec=\dicta\tilde\inputvec.
\end{align*}
Unfortunately, P0 has a prohibitive (combinatorial) computational complexity, even for small dimensions \inputdimA. 
One of the most popular and computationally tractable alternative to solving \Pzero is basis pursuit (BP)~\cite{chen1998,donoho2001,donoho2002,gribonval2003,elad2002,tropp2004}, which corresponds to the convex program
\begin{align*}
(\text{BP})  & \quad \mathit{minimize\;} 
\normone{\tilde\inputvec} \quad \mathit{subject\;to} \;  \outputvec=\dicta\tilde\inputvec.
\end{align*}
Recovery guarantees for P0 and BP are usually expressed in terms of the sparsity level $\nx=\normzero{\inputvec}$ and the coherence parameter of the dictionary~$\dicta$,  defined as
$\coha =\max_{k,\ell,k\neq \ell}  \,\abs{\veca^H_k\veca_\ell}$.
%
Specifically, a sufficient condition for \inputvec to be the unique solution of \Pzero and for BP to deliver this solution\footnote{The condition \fref{eq:classicalthreshold} also ensures perfect recovery using orthogonal matching pursuit~(OMP)~\cite{tropp2004,Pati1993,davis1994}, which is, however, not further investigated in this paper.} is~\cite{donoho2002,gribonval2003,tropp2004} 
\begin{align} \label{eq:classicalthreshold}
\nx<  \frac{1}{2}\!\left(1+\frac{1}{\coha}\right).
\end{align}
%

\subsection{Recovery of approximately sparse signals from noisy measurements}

For the case of bounded (otherwise arbitrary) measurement noise, i.e., \mbox{$\noiseout = \dicta\inputvec + \noise$} with $\normtwo{\noise}\leq\varepsilon$, recovery guarantees based on the coherence parameter \coha were developed in~\cite{donoho2006a,fuchs2006,tropp2006,benhaim2010,CWX10}.
The corresponding recovery conditions mostly treat the case of perfectly-sparse signals, i.e., where only a small fraction of the entries \inputvec are nonzero. 
\revision{Fortunately, many real-world signals exhibit the property that most of the signal's energy (e.g., in terms of the sum of absolute values) is concentrated in only a few entries. We refer to this class of signals as \emph{approximately sparse} in the remainder of the paper.}
For such signals, the support set associated to the best $\nx$-sparse approximation is defined as
\revision{\begin{align*}
 \widehat{\setX} = \supp_{\nx}\!(\inputvec) = \argmin_{\widetilde{\setX}\in\Sigma_{\nx}}\, \|{\inputvec-\inputvec_{\widetilde{\setX}}}\|,
\end{align*}
where the set $\Sigma_{\nx}$ contains all support sets of size \nx corresponding to perfectly \nx-sparse vectors having the same dimension as  \inputvec.}
%
A particular sub-class of approximately sparse signals is the set of \emph{compressible signals}, whose approximation error decreases according to a power law~\cite{baraniuk2008}.


The following theorem provides a sufficient condition for which  a suitably modified version of BP, known as BP denoising (BPDN)~\cite{chen1998}, stably recovers an approximately sparse vector $\inputvec$ from the noisy observation \noiseout.
\begin{thm}[BP denoising~{\cite[Thm.~2.1]{CWX10}}]\label{thm:CaiRecoveryExt}
Let $\noiseout=\dicta\inputvec+\noise$, \mbox{$\normtwo{\noise}\leq\varepsilon$}, and $\setX=\supp_{\nx}\!(\inputvec)$. If \fref{eq:classicalthreshold} is met, then the solution $\hat{\inputvec}$ of the convex program 
\begin{align*}
(\mathrm{BPDN}) \quad \text{minimize} \,\,\normone{\tilde\vecx} \quad \text{subject to}\; \normtwo{\noiseout-\dicta\tilde\inputvec}\leq \eta
\end{align*}
with $\varepsilon\leq\eta$ satisfies
\begin{align} \label{eq:CaiRecoveryExtError}
\normtwo{\inputvec - \hat{\inputvec}} \leq C_0 (\varepsilon+\eta) + C_1\normone{\vecx-\vecx_{\setX}},
\end{align}
where both (non-negative) constants $C_0$ and $C_1$ depend on \coha and \nx.
\end{thm}
\begin{IEEEproof}
\revision{The proof in \cite[Thm.~2.1]{CWX10} is given for perfectly sparse vectors only. Since some of the proofs presented in the remainder of the paper are developed for approximately sparse signals, we detail the relevant steps that generalize \cite[Thm.~2.1]{CWX10} in \fref{app:CaiRecoveryExt}.}
\end{IEEEproof}

We emphasize that \emph{perfect} recovery of \inputvec is, in general, impossible in the presence of bounded (but otherwise arbitrary) measurement noise \noise.
Hence, we consider \emph{stable recovery} instead, i.e., in a sense that the \elltwo-norm of the difference between the estimate $\hat{\inputvec}$ and the ground truth \inputvec is bounded from above by the \elltwo-norm of the noise $\normtwo{\noise}$ and the best $\nx$-sparse approximation as in \fref{eq:CaiRecoveryExtError}.
\revision{The constants $C_0$ and $C_1$ depend on the coherence parameter $\coha$ and on $\nx$, and increase as one approaches the limits of \fref{eq:classicalthreshold}. As an example, we obtain $C_0\approx2.59$ and $C_1\approx0.16$ for $\coha=0.01$ and $n_x=20$.}
\revision{Note that if $\varepsilon$ is known, one should set $\varepsilon=\eta$ to minimize the error~\fref{eq:CaiRecoveryExtError}.}
We furthermore note that \fref{thm:CaiRecoveryExt} generalizes the results for noiseless measurements and perfectly sparse signals in~\cite{donoho2002,gribonval2003,tropp2004} using BP (cf.~\fref{sec:noiselesscase}).
\revision{Specifically, for $\normtwo{\noise}=0$ and $\normone{\vecx-\vecx_{\setX}}=0$, BPDN with $\varepsilon=\eta=0$ corresponds to BP and \emph{perfectly} recovers~\inputvec whenever \fref{eq:classicalthreshold} is met.}
%

\subsection{Recovery guarantees from sparsely corrupted measurements}

A large number of restoration and separation problems occurring in practice can be formulated as sparse signal recovery from sparsely corrupted signals using the input-output relation \fref{eq:systemmodel}.
Special or related cases of the general model~\fref{eq:systemmodel} have been studied in~\cite{donoho1989,GN08,laska2009democracy,laska,candes2005decoding,carrillo2010,wright2010,nguyen2011,kuppinger2010a,SKPB10,XiLi11}.

\paragraph{Probabilistic recovery guarantees}

Recovery guarantees for the probabilistic setting (i.e., recovery of \inputvec is guaranteed with high probability) for random (sub-)Gaussian matrices, which are of particular interest for applications based on \revision{compressive sensing (CS)~\cite{donoho2006,candes2006c},} have been reported in~\cite{laska2009democracy,wright2010,laska,XiLi11}. Similar results for randomly sub-sampled unitary matrices \dicta have been developed in~\cite{nguyen2011}. 
The problem of sparse signal recovery from a nonlinear measurement process in the presence of impulse noise was considered in~\cite{carrillo2010}, and probabilistic results for signal detection based on \ellone-norm minimization in the presence of impulse noise was investigated in \cite{candes2005decoding}.
\revision{Another strain of probabilistic recovery guarantees has considered perfectly sparse signals from noiseless measurements with randomness on the location and values of the coefficient vectors~\cite{tropp2008,kuppinger2010a,PBS12}.}
\revision{In the remainder of the paper, we will focus on the deterministic setting exclusively.}

\paragraph{Deterministic recovery guarantees}

Recovery guarantees in the deterministic setting for noiseless measurements and signals being perfectly sparse, i.e., the model $\noiseout = \dicta\inputvec + \dictb\error$, have been studied in~\cite{donoho1989,GN08,DK08,kuppinger2010a,SKPB10,ASPB12}.
In~\cite{donoho1989}, it has been shown that when $\dicta$ is the discrete Fourier transform (DFT) matrix, $\dictb=\bI_\outputdim$ and when the support set of the interference \error is known, perfect recovery of \inputvec is possible if $2\nx n_e < \outputdim$, where $\nerr=\normzero{\error}$.
The case of \dicta and \dictb being arbitrary dictionaries (whereas \inputvec and \error are assumed to be perfectly sparse and for noiseless measurements) has been studied  for different cases of support-set knowledge in \cite{kuppinger2010a,SKPB10}.
\revision{There, deterministic recovery guarantees depending on the number of nonzero entries $\nx$ and $\nerr$ in \inputvec and \error, respectively,} and on the coherence parameters \coha and \cohb of \dicta and \dictb, as well as on the \emph{mutual coherence} between the dictionaries \dicta and \dictb, which is defined as
$\cohm = \max_{k,\ell} \, \abs{\veca^H_k\vecb_\ell}$.
%
A summary of the recovery guarantees presented in \cite{kuppinger2010a,SKPB10} (along with the novel recovery guarantees presented in the next section) is given in \fref{tab:recguaranteesummary}, where, for the sake of simplicity of exposition, we define the following function:
\begin{align*}
  f(u,v) = \pos{1-\coha\!\left(u-1\right)}\pos{1-\cohb\!\left(v-1\right)}.
\end{align*}

We emphasize that the results presented in \cite{kuppinger2010a,SKPB10} are for perfectly sparse and noiseless measurements only, and furthermore, that the algorithms and proof techniques cannot be adapted for the more general setting proposed in \fref{eq:systemmodel}.
In order to gain insight into the practically more relevant case of approximately sparse signals and noisy measurements, we next develop new restoration and separation algorithms for several different cases of support-set knowledge and provide corresponding recovery guarantees.  Our results complement those in \cite{kuppinger2010a,SKPB10} (cf.~\fref{tab:recguaranteesummary}).

\revision{The case of signal separation with more than two orthonormal bases has been studied in \cite{GN08}. Those results  have been derived for perfectly sparse signals from noiseless measurements; a generalization of the results shown next to more than two dictionaries is left for future work.}
\revision{Another set of theoretical results for sparsity-based signal separation has been derived in \cite{DK08,Donoho:2010uq,Kutyniok:2011fk}. Those results focus on the \emph{analysis} separation problem in general (possibly infinite-dimensional) frames, which aims at minimizing the number of non-zero entries of the analysis coefficients rather than the synthesis coefficients considered here (see \cite{Kutyniok:2011fk} for the details). 
The recovery conditions have been derived using a joint concentration measure and the so-called \emph{cluster coherence}, which enable the derivation of recovery conditions for the analysis separation problem that explicitly exploit the structure of particular pairs of frames (e.g., wavelets and curvelets).  
Coherence-based results for hybrid synthesis--analysis problems for pairs of general dictionaries were developed recently in \cite{ASPB12}.}


\begin{table}
 \begin{minipage}[c]{\columnwidth}    \centering
   \caption{Summary of deterministic recovery guarantees for perfectly/approximately sparse signals that are corrupted by interference in the absence/presence of measurement noise.}
   \vspace{0.2cm}
   \label{tab:recguaranteesummary}
   \begin{tabular}{lccc}
     \toprule[0.15em]  
     Support-set & \multirow{2}{*}{Recovery condition} & Perfectly sparse  & Approx.~sparse \\
     knowledge & &  and no noise &  and noise \\
     \midrule[0.1em]
      \inputvec and \error & $\nx\nerr\cohm^2\!<\!f(\nx,\nerr)$ & \cite[Thm.~3]{SKPB10}  & \fref{thm:directrestoration} \\ 
     \midrule
     \error  only & $2\nx\nerr\cohm^2\!<\!f(2\nx,\nerr)$ & \cite[Thms.~4 and 5]{SKPB10} & \fref{thm:BPRES}\\
     \midrule
     \multirow{2}{*}{None} &  \cite[Eq.~12]{kuppinger2010a}\footnote{The recovery condition is valid for BP and OMP; a less restrictive condition for P0 is given in \cite[Thm.~2]{kuppinger2010a}.}  & \cite[Thm.~3]{kuppinger2010a} & --- \\
     & Eq.~\ref{eq:BPSEPcondition} & --- & \fref{thm:BPSEP} \\
      \bottomrule[0.15em]
\end{tabular}
\end{minipage}
\end{table} 


\section{Main Results}
\label{sec:mainresults}

We now develop several computationally efficient methods for restoration or separation under the model \fref{eq:systemmodel} and derive corresponding recovery conditions that guarantee their stability. 
Our recovery guarantees depend on the $\elltwo$-norm of the noise vector and on the amount of knowledge on the dominant nonzero entries of the signal and noise vectors. 
Specifically, we consider the following three cases:
\begin{inparaenum}
\item[\emph{1) Direct restoration:}] The locations of the entries corresponding to the best $\nx$-sparse approximation of \inputvec and the support set of the (perfectly sparse) interference vector \error are known prior to recovery, 
\item[\emph{2) BP restoration:}] Only the support set of \error is known, 
\item[\emph{3) BP separation:}] No knowledge about \inputvec and \error is available, except for the fact that each vector exhibits an approximately an sparse representation in \dicta and \dictb, respectively. 
\end{inparaenum}
%

\subsection{Direct restoration: Support-set knowledge of \inputvec and \error}
\label{sec:directrecovery}

We start by addressing the case where the locations of the dominant entries (in terms of absolute value) of the approximately sparse vector \inputvec and the support set $\setE$ associated with the perfectly sparse interference vector \error are known prior to recovery. 
This scenario is relevant, for example, in the restoration of old phonograph records \cite{VF92,GR98}, where one wants to recover a  band-limited signal that is impaired by impulse noise, such as clicks and pops. The occupied frequency band of phonograph recordings is typically known prior to recovery.
In this case, one may assume $\dicta$ to be the inverse $M$-dimensional discrete cosine transform (DCT) matrix and $\dictb=\bI_\outputdim$. The locations of the clicks and pops, i.e., the support set $\setE=\supp(\error)$, can be determined prior to recovery using the techniques described in~\cite{GR98}, for example.

The restoration approach considered for this setup is as follows. 
\revision{Since \setE and $\setX=\supp_{\nx}\!(\inputvec)$ are both known prior to recovery, we start by projecting the noisy observation vector \noiseout onto the orthogonal complement of the range space spanned by $\dictb_\setE$, which eliminates the sparse interference.}
Concretely, we consider
\begin{align} \label{eq:DRprojection}
 \bR_\setE \vecz &= \bR_\setE \!\left(\dicta \inputvec+\dictb\error_\setE +\noise\right) = \bR_\setE \dicta \inputvec + \bR_\setE\noise,
\end{align}
where $\bR_\setE = \bI_{\outputdim} - \dictb_\setE \dictb^\dagger_\setE$ is the projector onto the orthogonal complement of the range space of $\dictb_\setE$, and we used the fact that $\bR_\setE\dictb\error_\setE=\mathbf{0}_{\outputdim\times1}$.
Next, one can separate \fref{eq:DRprojection} by exploiting the fact that \setX is known
\begin{align*} 
 \bR_\setE \vecz & = \bR_\setE \dicta (\inputvec_\setX + \inputvec_{\setX^c})  + \bR_\setE\noise  = \bR_\setE \dicta_\setX [\inputvec]_\setX +  \bR_\setE \dicta \inputvec_{\setX^c} + \bR_\setE\noise
\end{align*}
and, \revision{by assuming $\bR_\setE\dicta_\setX$ has full rank}, we can isolate the dominant entries~$[\inputvec]_\setX$ as 
\begin{align} \label{eq:DRresidualerrors}
 (\bR_\setE\dicta_X)^\dagger\bR_\setE \vecz & =  [\inputvec]_\setX +  (\bR_\setE\dicta_X)^\dagger\bR_\setE (\dicta \inputvec_{\setX^c} + \noise).
\end{align}
\revision{In the case where both vectors $\inputvec_{\setX^c}$ and $\noise$ are equal to zero, we obtain 
\begin{align} \label{eq:DRcoefficientestimate}
 (\bR_\setE\dicta_X)^\dagger\bR_\setE \vecz = [\inputvec]_\setX,
\end{align}
and therefore the entries of \inputvec contained in the support set $\setX$ are recovered perfectly  by this approach.}
Note that conjugate gradient methods (e.g.,~\cite{hestenes1952methods}) offer an efficient way of computing~\fref{eq:DRcoefficientestimate}.

The following theorem provides a sufficient condition for $(\bR_\setE\dicta_X)^\dagger\bR_\setE$ to exist and for which the vector~\inputvec can be restored stably from the noisy measurement \noiseout using the direct restoration (DR) procedure outlined above. 

\begin{thm}[Direct restoration] \label{thm:directrestoration}
Let $\noiseout=\dicta\inputvec+\dictb\error+\noise$ with  $\normtwo{\noise}\leq\varepsilon$, \error perfectly \nerr-sparse with support set \setE, and $\setX=\supp_{\nx}\!(\inputvec)$. Furthermore, assume that the support sets \setE and \setX are known prior to recovery. If
\begin{align} \label{eq:DRcondition}
\nx \nerr  \cohm^2 < f(\nx,\nerr),
\end{align}
then \revision{$\bR_\setE\dicta_\setX$ is full rank} and the vector $\hat{\inputvec}$ computed according to 
\begin{align*}
  (\mathrm{DR}) \quad  [\hat{\inputvec}]_\setX=(\bR_\setE\dicta_\setX)^\dagger\bR_\setE\noiseout, \quad [\hat{\inputvec}]_{\setX^c} = \bZero_{\abs{\setX^c}\times1}
\end{align*} 
with $\bR_\setE = \bI_\outputdim - \dictb_\setE\dictb_\setE^\dagger$ satisfies
\begin{align*} 
\normtwo{\inputvec - \hat{\inputvec}} \leq C_3\varepsilon + C_4\normone{\vecx-\vecx_{\setX}},
\end{align*}
where the (non-negative) constants $C_3$ and $C_4$ depend on the coherence parameters \coha, \cohb, and \cohm, and on the sparsity levels \nx and \nerr.
\end{thm}
\begin{IEEEproof}
The proof is given in \fref{app:directrestoration}.
\end{IEEEproof}

\revision{\fref{thm:directrestoration} and in particular \fref{eq:DRcondition} provides a sufficient condition for which DR enables the stable recovery of  \inputvec from \noiseout.}
Specifically, \fref{eq:DRcondition} states that for a given number of sparse corruptions \nerr, the smaller the coherence parameters \coha, \cohb, and \cohm, the more dominant entries of \inputvec can be recovered stably from~\noiseout. 
The case that guarantees the recovery of the largest number of dominant entries in \inputvec is when \dicta and \dictb are orthonormal bases (ONBs) ($\coha=\cohb=0$) that are maximally incoherent ($\cohm=1/\sqrt{\outputdim}$); \revision{this is, for example, the case for the Fourier--identity pair, leading to the recovery condition $\nx\nerr<\outputdim$.}

\revision{The recovery guarantee in \fref{thm:directrestoration} generalizes that in \cite[Thm.~3]{SKPB10} to approximately sparse signals and noisy measurements. 
Since \fref{eq:DRcondition} is identical to the condition \cite[Thm.~3]{SKPB10} (cf.~\fref{tab:recguaranteesummary}) we see that considering approximately sparse signals and (bounded) measurement noise does \emph{not} result in a more restrictive recovery condition.}
We finally note that the recovery condition in \fref{eq:DRcondition} was shown in \cite{SKPB10} to be tight for certain  signals in the case where~\dicta is the DFT matrix and $\dictb=\bI_\outputdim$.

\subsection{BP restoration: Support-set knowledge of \error only}

Next, we find conditions guaranteeing the stable recovery in the setting where the support set of the interference vector \error is known prior to recovery. 
A prominent application for this setting is the restoration of saturated signals~\cite{laska2009democracy,Adler2011}. Here, no knowledge on the locations of the dominant entries of \inputvec is required. The support set \setE of the sparse interference vector can, however, be easily identified by comparing the measured signal entries $[\noiseout]_i$, $i=1,\ldots,\outputdim$, to a saturation threshold. 
\revision{Further application examples for this setting include the removal of impulse noise~\cite{VF92,GR98,novak2010}, as well as sparsity-based in-painting and super-resolution~\cite{mallat2010,elad2001,bertalmio2000}.}

The recovery procedure for this case is as follows. 
Since \setE is known prior to recovery, we may recover the vector \inputvec by projecting the noisy observation vector \noiseout onto the orthogonal complement of the range space spanned by~$\dictb_\setE$ (cf.~\fref{sec:directrecovery}).
This projection eliminates the sparse noise and leaves us with a sparse signal recovery problem similar to that in \fref{thm:CaiRecoveryExt}.
In particular, we consider recovery from 
\begin{align} \label{eq:projectionapproach}
 \bR_\setE \vecz &= \bR_\setE \!\left(\dicta \inputvec+\dictb\error_\setE +\noise\right) = \bR_\setE \dicta \inputvec + \bR_\setE\noise,
\end{align}
\revision{where $\bR_\setE = \bI_{\outputdim} - \dictb_\setE \dictb^\dagger_\setE$.}
%
%
The following theorem provides a sufficient condition that guarantees the stable restoration of the vector \inputvec from \fref{eq:projectionapproach}.

\begin{thm}[BP restoration] \label{thm:BPRES}
Let $\noiseout=\dicta\inputvec+\dictb\error+\noise$ with $\normtwo{\noise}\leq\varepsilon$. Assume~$\error$ to be perfectly $\nerr$-sparse and $\setE = \supp(\error)$ to be known prior to recovery. Furthermore, let $\setX=\supp_{\nx}\!(\inputvec)$. If
\begin{align} \label{eq:BPREScondition}
2\nx\nerr\cohm^2 <  f(2\nx,\nerr),
\end{align}
then the result $\hat{\inputvec}$ of BP restoration 
\begin{align*}
(\mathrm{BP}\text{-}\mathrm{RES}) \quad \left\{
\begin{array}{ll}
\text{minimize} & \normone{\tilde\vecx} \\ 
\text{subject to} & \normtwo{\bR_\setE(\noiseout-\dicta\tilde\inputvec)}\leq \eta
\end{array}\right.
\end{align*}
with $\bR_\setE = \bI_\outputdim - \dictb_\setE\dictb_\setE^\dagger$ and $\varepsilon\leq\eta$ satisfies
\begin{align*} 
\normtwo{\inputvec - \hat{\inputvec}} \leq C_5(\varepsilon+\eta) + C_6\normone{\inputvec-\inputvec_{\setX} },
\end{align*}
where the (non-negative) constants $C_5$ and $C_6$ depend on the coherence parameters \coha, \cohb, and \cohm, and on the sparsity levels \nx and \nerr.
\end{thm}
\begin{IEEEproof}
The proof is given in \fref{app:BPRES}. 
\end{IEEEproof}

The inequality \fref{eq:BPREScondition} provides a sufficient condition on the number \nx of dominant entries of \inputvec for which BP-RES can stably recover  \inputvec from \noiseout. The condition depends on the coherence parameters \coha, \cohb, and \cohm, and  the number of sparse corruptions~\nerr.
\revision{As for the case of DR, the situation that guarantees the recovery of the largest number \nx of dominant coefficients in \inputvec, is when \dicta and \dictb are maximally incoherent ONBs.
In this situation, \fref{eq:BPREScondition} reduces to $2\nx\nerr<\outputdim$, which is two times more restrictive than that for DR (see \cite{SKPB10} for an extensive discussion on this factor-of-two penalty).}

The following observations are immediate consequences of \fref{thm:BPRES}:
\begin{itemize}
\item  \revision{If the vector \inputvec is perfectly $\nx$-sparse and for noiseless measurements, BP-RES using $\eta=0$ \emph{perfectly} recovers \inputvec if \fref{eq:BPREScondition} is met. 
Note that two restoration procedures have been developed for this particular setting in \cite[Thms.~4 and 5]{SKPB10}. Both methods enable perfect recovery under exactly the same conditions (cf.~\fref{tab:recguaranteesummary}).}
Hence, generalizing the recovery procedure to approximately sparse signals and measurement noise does not incur a penalty in terms of the recovery condition.
\item The restoration method in  \cite[Thm.~5]{SKPB10} requires  a column-normalization procedure to guarantee perfect recovery under the condition  \fref{eq:BPREScondition}. Since in this special case, BP-RES (with $\eta=0$) corresponds to BP,  \fref{thm:BPRES} implies that this normalization procedure is not necessary for guaranteeing perfect recovery under \fref{eq:BPREScondition}.
\revision{Note, however, that this observation does not apply to OMP-based recovery (see \cite{PB11} for the details).}
\end{itemize}
\revision{We finally note that \fref{eq:BPREScondition} has been shown in \cite{SKPB10} to be tight for certain signal and interference pairs in the case where \dicta is the DFT matrix and $\dictb=\bI_\outputdim$.}

\subsection{BP separation: No knowledge on the support sets}
 
We finally consider the case where no knowledge about the support sets of the approximately sparse vectors \inputvec and \error is available.
A typical application scenario is \revision{signal separation~\cite{elad2005,cai2009,Kutyniok:2011fk},} e.g., the decomposition of audio, image, or video signals into two or more distinct features, i.e., in a part that exhibits an approximately sparse representation in the dictionary \dicta and another part that exhibits an approximately sparse representation in \dictb. 
Decomposition then amounts to performing simultaneous recovery of \inputvec and \error from \mbox{$\noiseout = \dicta\inputvec + \dictb\error + \noise$}, followed by computation of the individual signal features according to $\dicta\inputvec$ and $\dictb\error$.
The main idea underlying this signal-separation approach involves rewriting~\eqref{eq:systemmodel} as
\begin{align} \label{eq:dictionarysplitting}
  \vecz = \dict \vecw + \noise,
\end{align}
where $\dict=[\,\dicta\,\,\dictb\,]$ is the concatenated dictionary of \dicta and \dictb  and the stacked vector $\vecw^T=[\,\inputvec^T\,\,\error^T\,]$. Signal separation now amounts to performing BPDN on \fref{eq:dictionarysplitting} for recovery of \vecw from \noiseout.

A straightforward way to arrive at a corresponding deterministic recovery guarantee for this problem is to consider~\dict as the new dictionary with the \emph{dictionary coherence} defined as 
\begin{align} \label{eq:dictionarycoherence}
\coh = \max_{i,j,i\neq j}\, \abs{\vecd^H_i\vecd_j} = \max\big\{\coha,\cohb,\cohm\big\}.
\end{align}
\revision{One can now use BPDN to recover \vecw from \fref{eq:dictionarysplitting} and invoke \fref{thm:CaiRecoveryExt} with the recovery condition in \fref{eq:classicalthreshold}, resulting in 
\begin{align} \label{eq:straightforwardcondition}
 n_w <\frac{1}{2}\!\left(1+\frac{1}{\coh}\right)
\end{align}
with $n_w=n_x+n_e$.}
It is, however, important to realize that \fref{eq:straightforwardcondition} ignores the structure underlying the dictionary \dict, i.e., it does not take into account the fact that \dict is a concatenation of two dictionaries that are characterized by the coherence parameters \coha, \cohb, and \cohm. Hence, the \revision{recovery condition~\fref{eq:straightforwardcondition} does not provide insight into which pairs of dictionaries \dicta and \dictb are most useful for signal separation.}
\revision{The following theorem improves upon \fref{eq:straightforwardcondition} by taking into account the structure underlying~\dict and enables us to gain insight into which pairs of dictionaries enable signal separation.}

\begin{thm}[BP separation] \label{thm:BPSEP}
Let $\noiseout=\dict \vecw+\noise$, with $\dict=[\,\dicta\,\,\dictb\,]$,  $\vecw^T = [\,\vecx^T\,\,\vece^T\,]$, and \mbox{$\normtwo{\noise}\leq\varepsilon$}. The dictionary \dict is characterized by the coherence parameters \coha, \cohb, \cohm, and \coh, and we assume $\cohb\leq\coha$ without loss of generality. 
\revision{If
\begin{align} \label{eq:BPSEPcondition}
  n_w = n_x+n_e < \max\left\{\frac{2(1+\coha)}{\coha+2\coh+\sqrt{\coha^2+\cohm^2}}\,,\,\frac{1+\coh}{2\coh} \right\},
\end{align}
then the solution $\hat{\vecw}$ of BP separation
\begin{align*}
(\mathrm{BP}\text{-}\mathrm{SEP}) \quad \left\{\begin{array}{ll}
\text{minimize} &  \normone{\tilde\vecw}  \\
\text{subject to} & \normtwo{\noiseout-\dict\tilde\vecw}\leq \eta
\end{array}\right.
\end{align*}
using $\varepsilon\leq\eta$  satisfies
\begin{align}
\normtwo{\vecw - \hat{\vecw}} \leq C_7 (\varepsilon+\eta) + C_8 \normone{\vecw-\vecw_\setW},
\end{align}
with $\setW=\supp_{n_w}(\vecw)$ and the (non-negative) constants $C_7$ and $C_8$.}
\end{thm}
\begin{IEEEproof}
The proof is given in \fref{app:BPSEP}.
\end{IEEEproof}

\revision{The recovery condition \fref{eq:BPSEPcondition} refines that in \fref{eq:straightforwardcondition}. Consider the two-ONB setting for which $\coha=\cohb=0$ and $\cohm=\coh$.
In this case, \fref{eq:dictionarycoherence} corresponds to $n_w=n_x+n_e<(1+1/\coh)/2$, whereas the condition for BP separation \fref{eq:BPSEPcondition} is given by
\begin{align} \label{eq:BPSEPtwoonbcondition}
 n_w=n_x+n_e< \frac{2}{3\coh}.
\end{align}
Hence, \fref{eq:BPSEPcondition} guarantees the stable recovery for a larger number of dominant entries in the stacked vector $\vecw^T=[\,\vecx^T\,\vece^T\,]$.
Recovery guarantees for perfectly sparse signals and noiseless measurements in the case of two ONBs have been developed in~\cite{elad2002,feuer2003,gribonval2003,GN08}.
The corresponding recovery condition $n_w=n_x+n_e < (\sqrt{2}-0.5)/\coh$ turns out to be less restrictive than the recovery condition for approximately sparse signals and measurement noise in~\fref{eq:BPSEPtwoonbcondition}.
Whether this behavior is a fundamental result of  considering approximately sparse signals and noisy measurements or is an artifact of the proof technique is part of on-going work.}


\section{Coherence-based Bounds on Restricted Isometry Constants}
\label{sec:discussion}

\revision{Coherence-based bounds on restricted isometry constants (RICs) are useful to efficiently compute bounds on the RIC that would otherwise require a combinatorial search~\cite{Pfetsch2012}.
Moreover, such bounds enable us to develop an alternative set of recovery conditions from the restricted isometry property (RIP) framework~\cite{candes2005decoding,CRT06,C08,CXZ09,CWX10a,CWX10b,CZ12}.
We next show RIC bounds all three cases of support-set knowledge and provide a comparison with the recovery guarantees obtained in the previous section, those from the RIP framework, and with existing ones from~\cite{kuppinger2010a,SKPB10} (recall~\fref{tab:recguaranteesummary}).}

\subsection{Restricted isometry property (RIP)}

\revision{An alternative route of obtaining deterministic recovery guarantees for approximately sparse signals and measurement noise, i.e., for \mbox{$\noiseout=\dict\inputvec+\noise$}, has been developed under the RIP framework~\cite{candes2005decoding,CRT06,C08,CXZ09,CWX10a,CWX10b,CZ12}.}
There, the dictionary~\dict is characterized by RICs rather than the coherence parameter $\mu_d$.
\begin{defi}[Restricted isometry constant (RIC) \cite{candes2005decoding}]
For each integer $\nx\geq1$, the RIC $\RIC{\nx}$ of~$\dict$ is the smallest number such that 
\begin{align} \label{eq:standardRICa}
\left(1-\RIC{\nx}\right)\normtwo{\inputvec}^2 \leq \normtwo{\dict\inputvec}^2 \leq \left(1+\RIC{\nx}\right)\normtwo{\inputvec}^2
\end{align}
holds for all perfectly $\nx$-sparse vectors $\inputvec$. 
\end{defi}

\revision{Stable recovery of~\inputvec with $\normtwo{\noise}\leq\varepsilon$ using BDPN can be guaranteed if the dictionary $\dict$ satisfies a restricted isometry property (RIP), e.g., of the form (i) \mbox{$\delta_{2\nx}<\sqrt{2}-1$} \cite{C08} or (ii)  $\delta_{\nx}<1/3$~\cite{CZ12}.
The main issue with such recovery conditions is the fact that computation of the RIC requires a combinatorial search~\cite{Pfetsch2012}.
Nevertheless, it has been shown in \cite{candes2008a,baraniuk2008a} that random dictionaries $\dict$ (e.g., with i.i.d.\ (sub)-Gaussian entries) satisfy such RIP conditions with high probability, which is of particular interest in CS~\cite{donoho2006,candes2006c}.}

\revision{
To arrive at recovery conditions that are explicit in the number of nonzero entries \nx and can be computed efficiently, one may bound the RIC in~\fref{eq:standardRICa} using the coherence parameter $\mu_d$ as $\RIC{\nx}\leq \mu_d(\nx-1)$~\cite{CXZ09,CWX10}.
Such bounds in combination with the RIP conditions in \cite{C08,CZ12}  can be used to derive alternative recovery conditions (i) $\nx<(1+(\sqrt{2}-1)/\mu_d)/2$ or (ii) $\nx<1+1/(3\mu_d)$, which are, in general, more restrictive than  \fref{eq:classicalthreshold}.}

\subsection{Coherence-based RIC bounds for sparsely corrupted signals}

\revision{We next provide coherence-based bounds on the RIC for DR, BP-RES, and BP-SEP, and derive corresponding  alternative recovery guarantees using results obtained in the RIP framework~\cite{C08,CWX10b}.}
%

\paragraph{RIC bound for signal restoration} \label{sec:RICBPRES}
As a byproduct of the proof for BP-RES detailed in \fref{app:BPRES}, the following coherence-based upper bound on the RIC for the matrix $\widetilde{\dicta}=\bR_\setE\dicta$ has been obtained:
\begin{lem}[{RIC bound for $\widetilde{\dicta}$}] \label{lem:RICboundEknown}
Let $\widetilde{\dicta}=\bR_\setE\dicta$ with $\bR_\setE = \bI_{\outputdim} - \dictb_\setE \dictb^\dagger_\setE$. For each integer $\nx\geq1$, the smallest number $\delta_{\nx}$ such that 
\begin{align*}
\left(1-\delta_{\nx}\right)\normtwo{\inputvec}^2 \leq \normtwo{\widetilde{\dicta}\inputvec}^2 \leq \left(1+\RIC{\nx}\right)\normtwo{\inputvec}^2
\end{align*}
holds for all perfectly $\nx$-sparse vectors $\inputvec\in\complexset^{N_a}$, is bounded from above by
\begin{align} \label{eq:BPRESRICbound}
\delta_{\nx} \leq \coha(\nx-1) + \frac{\nx\nerr \cohm^2}{\pos{1-\cohb(\nerr-1)}}.
\end{align}
\end{lem}

For direct recovery (DR), a recovery condition from the RIP framework follows straightforwardly from \fref{lem:RICboundEknown}; i.e., we need $\delta_{n_x}<1$ to ensure that the inverse of $\widetilde{\dicta}=\bR_\setE\dicta$ exists and to enable the stable recovery of $\vecx$ using $(\text{DR})$. 
\revision{For BP restoration, a recovery condition from the RIP framework is obtained by combining  \fref{eq:BPRESRICbound} with the RIP condition $\delta_{\nx}<1/3$ from~\cite{CZ12}.
We note that for \mbox{$\mu_d<1/3$} this condition is more restrictive than the  condition~\fref{eq:BPREScondition}.}
\revision{Hence, for most relevant values of $\mu_d$, the recovery condition in~\fref{eq:BPREScondition} is less restrictive than those obtained from the RIP framework.}

\paragraph{RIC bound for signal separation}
\revision{As a byproduct of the proof for BP-SEP detailed in \fref{app:BPSEP}, the following coherence-based upper bound on the RIC for the concatenated dictionary $\dict=[\,\dicta\,\,\dictb\,]$ has been obtained:}
\begin{lem}[{RIC bound for $\dict$}] \label{cor:RICboundD}
 Let $\dict=[\,\dicta\,\,\dictb\,]$ be characterized by \coha, \cohb, \cohm, and \coh, and assume $\cohb\leq\coha$ without loss of generality. For each integer $n_w\geq1$ the smallest number $\RIC{n_w}$ such that 
\begin{align*}
\left(1-\RIC{n_w}\right)\normtwo{\vecw}^2 \leq \normtwo{\dict\vecw}^2 \leq \left(1+\RIC{n_w}\right)\normtwo{\vecw}^2
\end{align*}
holds for all perfectly $n_w$-sparse vectors $\vecw\in\complexset^{N_a+N_b}$, is bounded by
\begin{align} \label{eq:RICboundD}
  \RIC{n_w} \leq  \min\left\{\frac{1}{2}\!\left(\coha(n_w-2)+n_w\sqrt{\coha^2+\cohm^2}\right),\,\coh(n_w-1) \right\}.
\end{align}
\end{lem}

\revision{As shown above, one can use the right hand side (RHS) of \fref{eq:RICboundD} in combination with RIP  condition $\delta_{n_w}<1/3$ of~\cite{CZ12} to obtain an alternative recovery guarantee for signal separation. 
As for BP restoration, this condition is more restrictive than that in \fref{eq:BPSEPcondition} in most cases.
Surprisingly, in the two-ONB case (i.e., $\coha=\cohb=0$ and $\cohm=\coh$), the recovery condition for signal separation obtained from the RIP framework coincides to that in~\fref{eq:BPSEPtwoonbcondition}.}

\subsection{Comparison of the Recovery Guarantees}

\begin{figure}[t]
\centering
 \includegraphics[width=0.85\columnwidth]{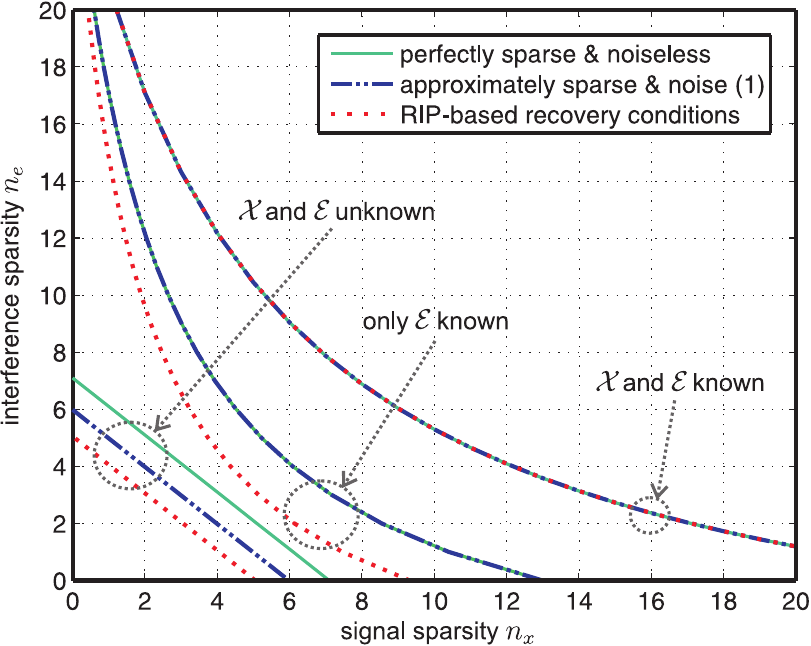}
  \caption{Comparison of the recovery conditions for $\cohm=\coh=0.1$ and $\coha=\cohb=0.04$. \revision{For the case ``$\setX$ and $\setE$ known,'' all three recovery conditions coincide; for the case ``only $\setE$ known,'' the conditions for perfectly and approximately sparse signals coincide.} }
  \label{fig:thresholds}
\end{figure}

\revision{Figure~\ref{fig:thresholds} compares the recovery conditions for the general model \fref{eq:systemmodel} to those obtained in \cite{kuppinger2010a,SKPB10} for perfectly sparse signals and noiseless measurements (see also \fref{tab:recguaranteesummary}), and the conditions derived from the RIP framework.
We compare the three different cases analyzed in \fref{sec:mainresults}:}
\begin{itemize}
\item \emph{Direct restoration:}
\revision{For DR, the recovery conditions for the general model \fref{eq:systemmodel} detailed in \fref{eq:DRcondition}, the condition in \cite[Eq.~11]{SKPB10} for perfectly sparse signals and noiseless measurements, and those obtained through the RIP framework coincide. Hence, the generalization  to approximately sparse signals and measurement noise does not incur a degradation in terms of the recovery condition.}

\item \emph{BP restoration:}
The recovery conditions for the general setup considered in this paper and the condition \cite[Eq.~14]{SKPB10} for perfectly sparse signals and noiseless measurements also coincide. \revision{Again, generalizing the results does not incur a loss in terms of the recovery conditions.}
As expected, the recovery condition obtained trough the RIP framework turns out to be  more restrictive (cf.~\fref{sec:RICBPRES}). 

\item \emph{BP separation:} 
\revision{We see that all of the recovery conditions differ. In particular, the condition \cite[Eq.~13]{kuppinger2010a} for  perfectly sparse signals and noiseless measurements is less restrictive than~\fref{eq:systemmodel}. As expected, the recovery condition from the RIP framework is most restrictive.}
\end{itemize}

In summary, we  see that having more knowledge on the support sets prior to recovery yields less restrictive recovery conditions. This intuitive behavior can also be observed in practice and is illustrated in \fref{sec:application}. 

\revision{We finally emphasize that all of the recovery conditions derived above are deterministic in nature and therefore conservative in the sense that, in practice, recovery often succeeds for sparsity levels \nx and \nerr much higher than the corresponding guarantees indicate. 
In particular, it is well-known that coherence-based deterministic recovery guarantees are typically limited by the so-called \emph{square-root bottleneck}, e.g.,~\cite{tropp2008,kuppinger2010a,SKPB10,PBS12}, as they are valid for \emph{all} dictionary pairs \dicta and \dictb with given coherence parameters, and all signal and interference realizations with given sparsity levels \nx and \nerr.
Nevertheless, we next show that our recovery conditions enable us to gain considerable insights into practical applications; i.e., they are useful for identification of appropriate dictionary pairs that should be used for sparsity-based signal restoration or separation.}

%
\section{Application Examples}
\label{sec:application}

We now develop two application examples to illustrate the main results of the paper. First, we show that direct restoration, BP restoration, and BP separation can be used for simultaneous denoising and declicking of corrupted speech signals. Then, we illustrate the impact of support-set knowledge for a sparsity-based in-painting example. 

\subsection{Simultaneous denoising and declicking}

In this example, we attempt the recovery of a speech signal that has been corrupted by a combination of additive Gaussian noise and impulse noise.
The main goal of this example is to illustrate the performance of our algorithms and not to benchmark the performance relative to existing methods;  a detailed performance and restoration-complexity comparison with existing methods for simultaneous denosing and declicking  is left for future work.

We corrupt a 9.5\,s segment (44\,100\,kHz sampling rate and 16\,bit precision) from the speech signal in~\cite{RB07} by adding zero-mean i.i.d.\ Gaussian noise and impulse noise.
\revision{The amplitudes of the audio signal has been normalized to the range $[-1,1]$.}
The variance of the additive noise is chosen such that the \revision{signal-to-noise ratio (SNR)} between the $L$-dimensional original audio signal~$\outputvec$ and the noisy version $\tilde{\outputvec}$,  defined as
\begin{align*}
\textsf{SNR} = 10\log_{10}\!\left({\normtwo{\outputvec}^2}/{\normtwo{\outputvec-\tilde{\outputvec}}^2}\right), 
\end{align*}
is \revision{$10.7$\,dB.}
The impulse interference (used to model the clicking artifacts in the audio signal) is generated as follows: We corrupt 10\% of the samples and chose the locations of the random clicks, which are modeled by the interference vector \error, uniformly at random. We then generate the clicks at these locations by adding i.i.d.\ zero-mean Gaussian random samples with variance $0.1$ to the noisy signal.
\revision{The resulting SNR is $0.31$\,dB.}

\paragraph{Recovery procedure}

Recovery is performed with overlapping blocks of dimension $\outputdim=1024$. The amount of overlap between adjacent blocks is \revision{128 samples.} 
We set~\dicta to the $1024\times1024$  DCT matrix, $\dictb=\bI_\outputdim$, and perform recovery based on $\noiseout=\dicta\inputvec+\error+\noise$.
The main reasons for using the DCT matrix in this example are
\begin{inparaenum}[(i)]
\item the speech signal is approximately sparse in the DCT basis;
\item \revision{we have $\coha=\cohb=0$; and}
\item the mutual coherence of the DCT--identity pair is small, i.e., \revision{$\cohm \approx 0.0442$,} which leads to less restrictive recovery conditions \fref{eq:DRcondition}, \fref{eq:BPREScondition}, and \fref{eq:BPSEPcondition}.
\end{inparaenum}
For all three recovery methods, we first compute an estimate $\hat{\inputvec}$ of \inputvec (and of \error in the case of BP separation) followed by computing an estimate of speech signal according to $\hat{\outputvec}=\dicta\hat{\inputvec}$. In order to reduce undesired artifacts occurring at the boundaries between two adjacent blocks, we overlap and add the recovered blocks using a raised-cosine window function when re-synthesizing the entire speech signal.

\begin{figure}[tp]
\centering
\subfigure[Original signal]{
\includegraphics[width=0.475\columnwidth]{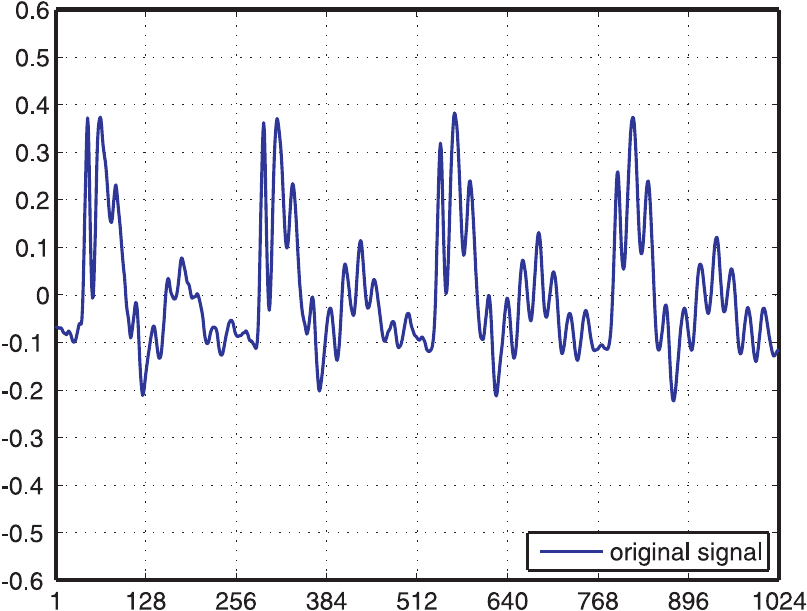}
\label{fig:bpres_fig1}
}
\subfigure[\revision{Noisy signal ($\textsf{SNR}=10.7$\,dB)}]{
\includegraphics[width=0.475\columnwidth]{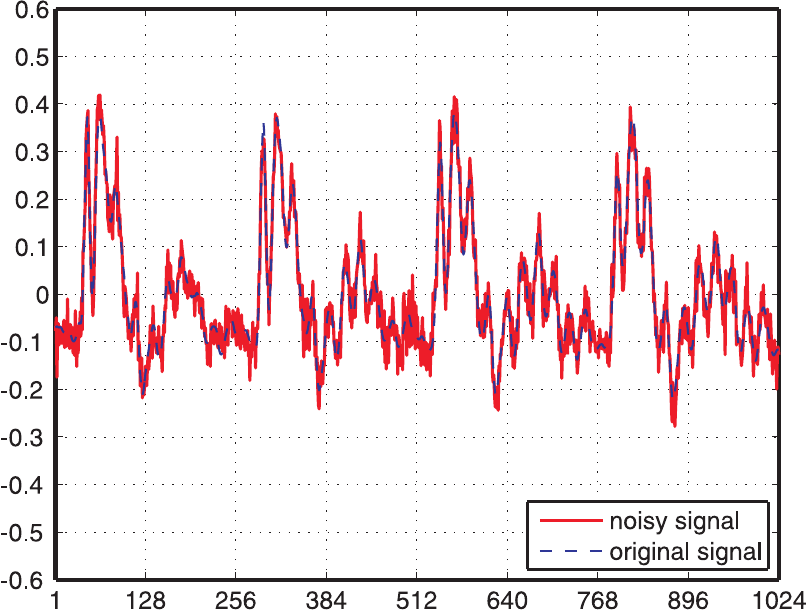}
\label{fig:bpres_fig2}
}

\subfigure[\revision{Noisy with clicks ($\textsf{SNR}=0.31$\,dB)}]{
\includegraphics[width=0.475\columnwidth]{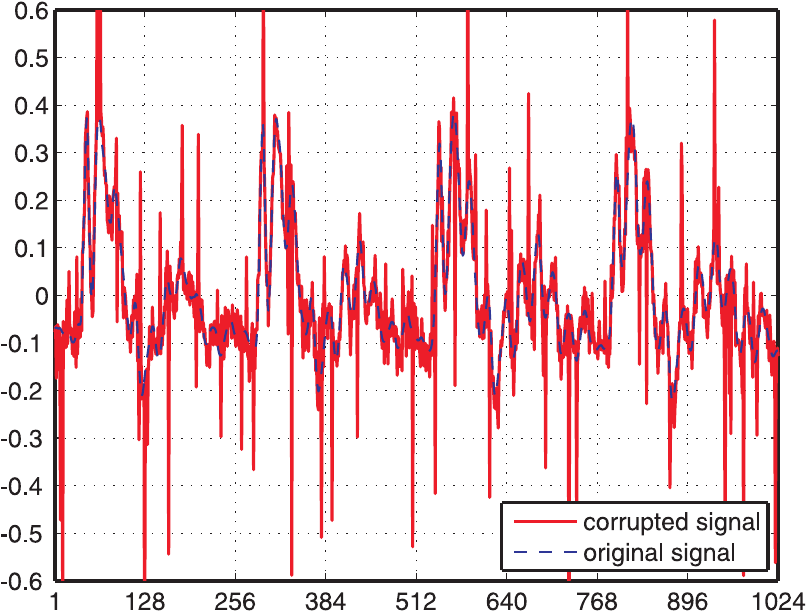}
\label{fig:bpres_fig3}
}
\subfigure[\revision{Direct restoration ($\textsf{SNR}=17.4$\,dB)}]{
\includegraphics[width=0.475\columnwidth]{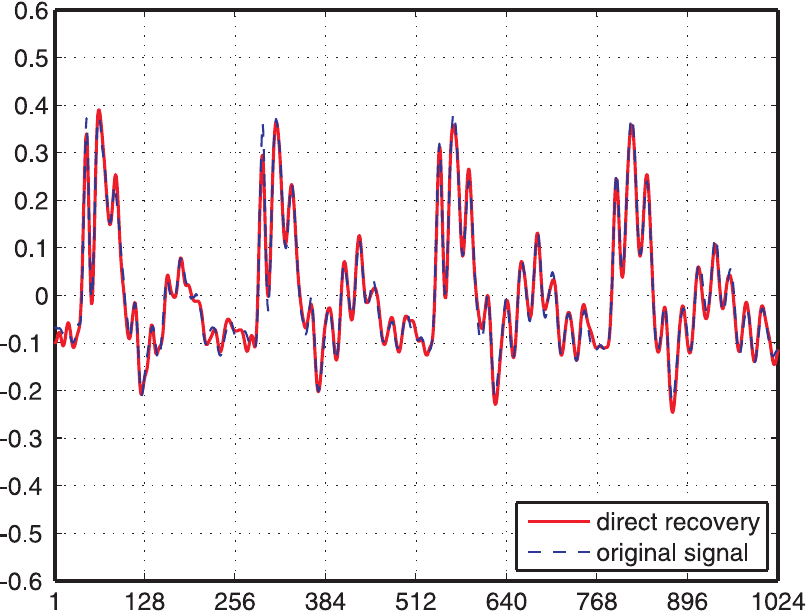}
\label{fig:bpres_fig5}
}

\subfigure[\revision{BP restoration ($\textsf{SNR}=15.5$\,dB)}]{
\includegraphics[width=0.475\columnwidth]{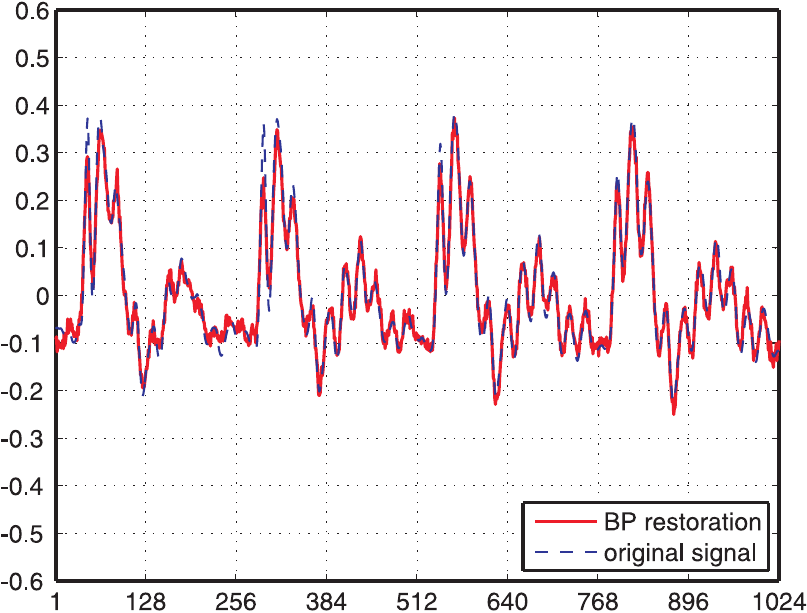}
\label{fig:bpres_fig4}
}
\subfigure[\revision{BP separation ($\textsf{SNR}=13.0$\,dB)}]{
\includegraphics[width=0.475\columnwidth]{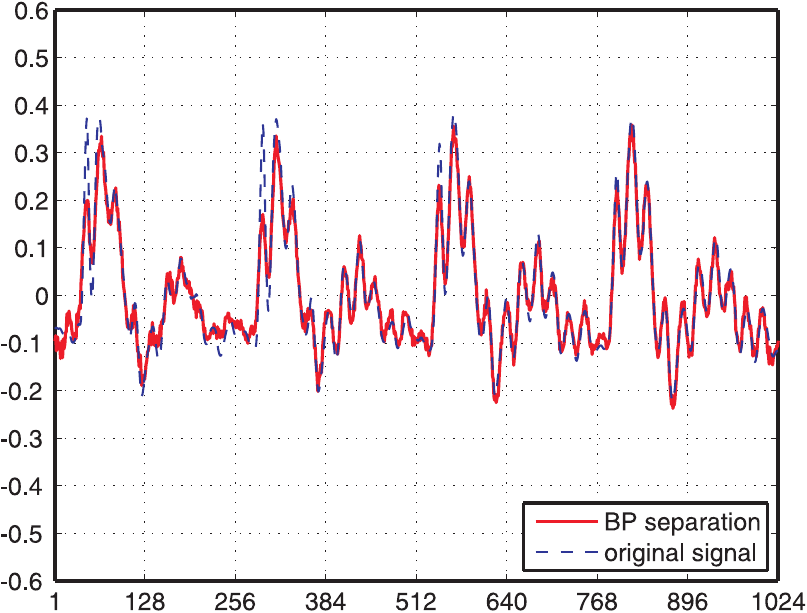}
\label{fig:bpres_fig6}
}

\caption{\revision{Signal-to-noise ratio (SNR) results of simultaneous reduction of Gaussian noise and interference in a corrupted speech signal (the x-axes correspond to sample indices, the y-axes to magnitudes).}} 
\label{fig:bpres_fig}
\end{figure}

\paragraph{Recovery results}

Figure~\ref{fig:bpres_fig} shows snapshots of the corruption and recovery procedure and the associated SNR values. The individual results of the three recovery procedures analyzed in this paper are as follows:
\begin{itemize}
\item \emph{Direct restoration:} In this case, the locations \setE of the impulse noise realizations are assumed to be known prior to recovery. 
\revision{To this end, we compute the DCT coefficients of the uncorrupted signal to identify the 128 largest (in magnitude) coefficients in each block.
This genie-aided support-set estimate is then used in the DR recovery stage. 
The SNR of the signal recovered through DR is $17.4$\,dB and, hence, DR is able to  improve the SNR by roughly $17$\,dB (see \fref{fig:bpres_fig5}). }
\item \emph{BP restoration:} In this case, the locations of the impulse noise spikes \setE are assumed to be known prior to recovery, but nothing is known about~\inputvec. \revision{We perform BP restoration with $\eta=0.8$, which results in an SNR of $15.5$\,dB (see \fref{fig:bpres_fig4}). Note that the parameter~$\eta$ determines the amount of denoising (for the Gaussian noise) and is used to optimize the resulting SNR.}
\item \emph{BP separation:}  In this case, we assume that nothing is known about the support sets of either \inputvec or \error. \revision{We perform BP-SEP with $\eta=0.8$ and discard the recovered error component~\error; the resulting SNR corresponds to $13.0$\,dB (see \fref{fig:bpres_fig6}).}
BP separation achieves surprisingly good recovery performance (compared to DR and BP-RES), while being completely \emph{blind} to the locations of the \revision{sparse interference. Hence, BP separation offers an elegant way to mitigate impulse noise in speech signals, without requiring sophisticated algorithms that detect the locations of the sparse interference (e.g., clicks and pops).}
\end{itemize}

\paragraph{Alternative recovery procedure}

\revision{
Rectangular windowing, as used in the example above, is known to yield sub-optimal sparsification of audio signals in the DCT basis. 
Furthermore, over-complete dictionaries often enable sparser representations than orthonormal bases. 
Therefore, it is natural to ask what happens if recovery is performed \emph{directly} on windowed audio signals using a redundant dictionary. 
To answer this question, we alternatively perform signal restoration using BP-RES\footnote{The performance for DR and BP-SEP behaves analogously to that of BP-RES.} on the basis of $\bW\vecz=\bW(\dicta\vecx+\dictb\vecn+\noise)$ rather than on~\fref{eq:systemmodel}, with $\bW$ denoting a diagonal matrix containing the windowing coefficients on the main diagonal. Moreover, we set $\bA$ to a redundant (or over-complete) DCT matrix~\cite{eladbook2010}.
To perform windowing within BP-RES, we use $\tilde{\vecz}=\bW\vecz$, $\widetilde{\bA}=\bW\bA$, and $\widetilde{\bB}=\bW\bB$, instead of $\vecz$, $\bA$, and $\bB$.\footnote{The dictionaries $\widetilde{\bA}$ and $\widetilde{\bB}$ were  normalized to have unit-norm columns. Note that re-normalizing the windowed identity basis $\widetilde{\bB}$ leads to $\bI_M$ and, hence, we have $\cohb=0$.}}

\revision{
\fref{tab:audiorecovery} shows the coherence parameters of $\bA$ and $\widetilde{\bA}$, the mutual coherences $\cohm$, and the recovery SNR for windowing after and within BP-RES, and for different redundancies. 
We can see that the SNR for the windowing approach within BP-RES is worse than that of the approach used above (cf.~\fref{fig:bpres_fig}). The reason for this behavior is the fact that even if windowing of~$\bA$ improves sparsification of $\bmx$, it also increases the coherence parameter~$\coha$. 
The recovery condition \fref{eq:BPREScondition} reflects this behavior and shows that even for small coherence values, a strong sparsification is necessary. 
\fref{tab:audiorecovery} furthermore shows that increasing the redundancy of $\dicta$ also degrades the SNR. This behavior is, once again, a result of the fact that the coherence of $\bA$ (or the windowed version $\widetilde{\bA}$) increases with the redundancy.
We conclude that windowing within BP-RES and/or increasing the redundancy in $\bA$ does not improve the performance in the considered audio-restoration example. 
}

\begin{table}[tb]
\centering
\caption{Coherence parameters and SNR of BP-RES for two windowing methods and redundant DCT dictionaries $\bA$ for $M=N_b=1024$ and a $2\times128$ raised-cosine window.}
\vspace{0.2cm}
\begin{tabular}{c|ccc|ccc}
\toprule[0.15em]
& \multicolumn{3}{|c|}{windowing after BP-RES} & \multicolumn{3}{|c}{windowing within BP-RES} \\
$N_a$ & $\coha$ & $\cohm$ & \textsf{SNR} & $\coha$ & $\cohm$ & \textsf{SNR} \\
\midrule[0.1em]
1024 & 0 & 0.0442 & 15.5         & 0.3824 & 0.0522 & 13.3  \\
1152 & 0.0447 & 0.0447 & 14.9 & 0.4062 & 0.0523 & 12.9 \\
1280 & 0.0658 & 0.0449 & 14.6 & 0.4127 & 0.0523 & 12.7 \\
1536 & 0.0854 & 0.0452 & 14.3 & 0.3737 & 0.0521 & 12.5 \\
2048 & 0.1147 & 0.0456 & 13.8 & 0.4185 & 0.0526 & 12.1 \\
\bottomrule[0.15em]
\end{tabular}
\label{tab:audiorecovery}
\end{table}

\paragraph{Discussion of the results}

The results shown above show that more knowledge on the support sets \setX and/or \setE leads to improved recovery results (i.e., larger SNR).
\revision{We emphasize that DR, BP restoration, and BP separation are all able to \emph{simulatenously} reduce Gaussian noise and impulse interference as the resulting SNR values are all larger than $10.7$\,dB (corresponding to the SNR of the noisy signal). The recovery procedure one should use in practice depends on the amount of support-set knowledge available prior to recovery.}

\revision{The use of redundant dictionaries or windowing within the restoration procedure did not show any advantage (over rectangular windowing and the use of ONBs) in the considered example. Nevertheless, we believe that specifically trained (or learned) dictionaries, e.g., using the method in~\cite{aharon2006}, have the potential to further improve the recovery performance; the exploration of methods that also maintain incoherence to other dictionaries is part of on-going work.}

We furthermore note that noise and clicks removal in audio signals is a well-studied topic in the literature (see, e.g.,  \cite{GR98} and references therein). 
\revision{However, most of the established methods rely on Bayesian estimation techniques, e.g., \cite{GR98,FG06,GCFW07}, which do not have theoretical guarantees. Sparsity-based  audio restoration  has been proposed recently in~\cite{Adler2011,AEJEGP11}; however, no recovery guarantees are available for the associated restoration algorithms.}

We finally emphasize that virtually all proposed methods require knowledge of the locations of the sparse corruptions prior to recovery, whereas our results for BP separation show that sparse errors can effectively be removed \emph{blindly} from speech signals.

\subsection{Removal of scratches in color photographs}

We now consider a simple sparsity-based in-painting application. While a plethora of in-painting methods have been proposed in the literature (see, e.g.,~\cite{bertalmio2000,WO08,CCSS09,cai2009} and the references therein), our goal here is to not to benchmark our performance vs.\ theirs but rather to illucidate the differences between BP restoration and BP separation, i.e., \revision{to quantify the impact of support-set knowledge and of the coherence parameters on the inpainting performance.}

\begin{figure}[tp]
\centering
\subfigure[Original]{
\includegraphics[width=0.223\columnwidth]{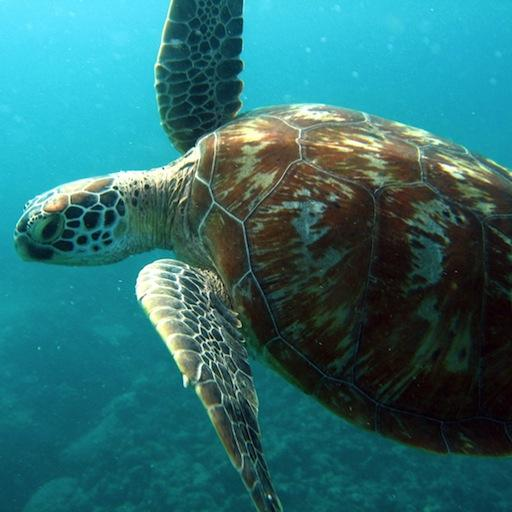}
\label{fig:scratch1}
}
\subfigure[Corrupted]{
\includegraphics[width=0.223\columnwidth]{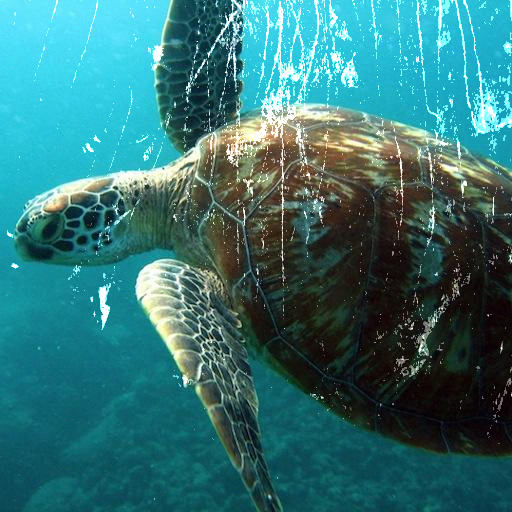}
\label{fig:scratch2}
}
\subfigure[R: DCT--identity]{
\includegraphics[width=0.223\columnwidth]{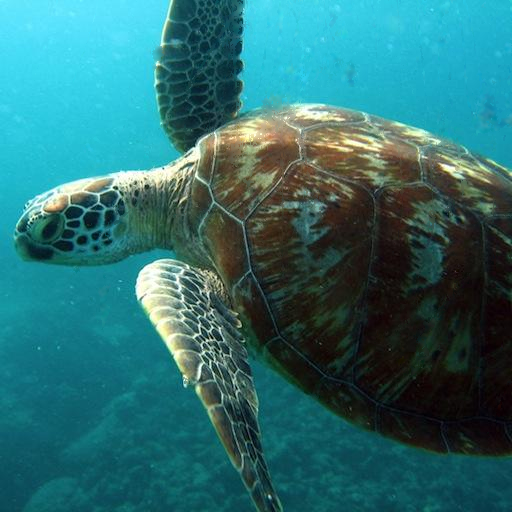}
\label{fig:scratch3}
}
\subfigure[S: DCT--identity]{
\includegraphics[width=0.223\columnwidth]{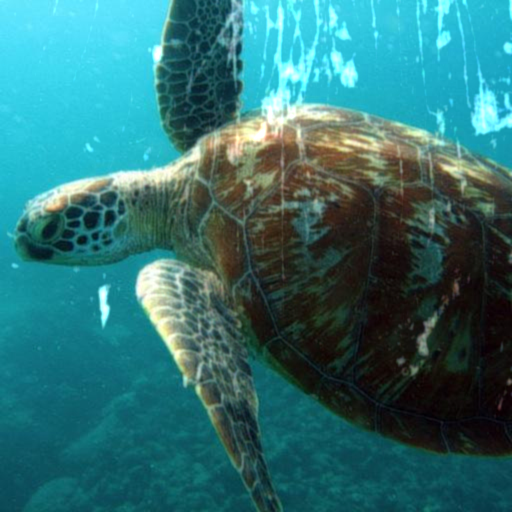}
\label{fig:scratch4}
}

\subfigure[R: DCT--DWT]{
\includegraphics[width=0.223\columnwidth]{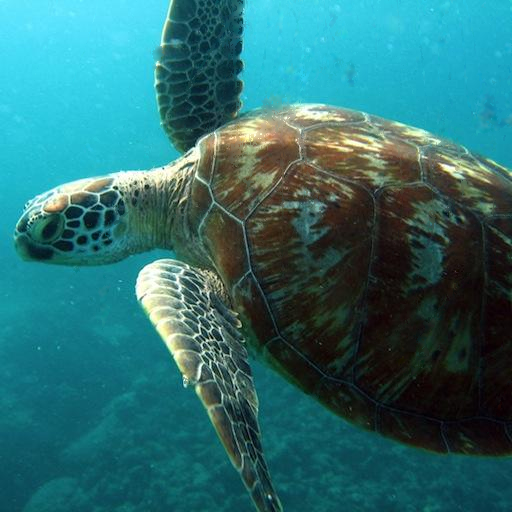}
\label{fig:scratch5}
}
\subfigure[S: DCT--DWT]{
\includegraphics[width=0.223\columnwidth]{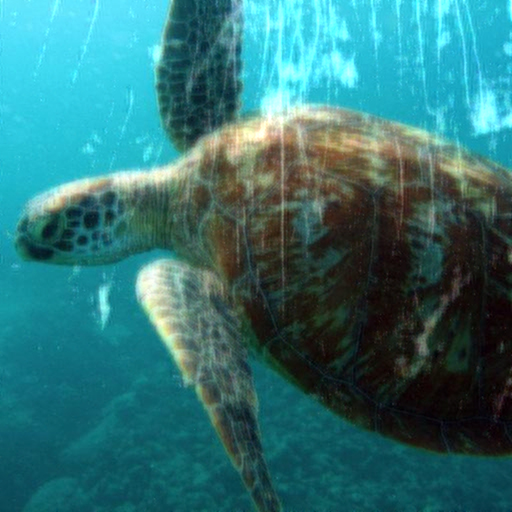}
\label{fig:scratch6}
}
\subfigure[R: DWT--identity]{
\includegraphics[width=0.223\columnwidth]{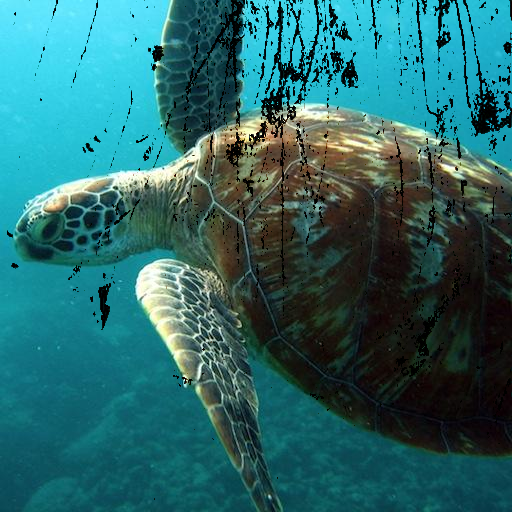}
\label{fig:scratch7}
}
\subfigure[S: DWT--identity]{
\includegraphics[width=0.223\columnwidth]{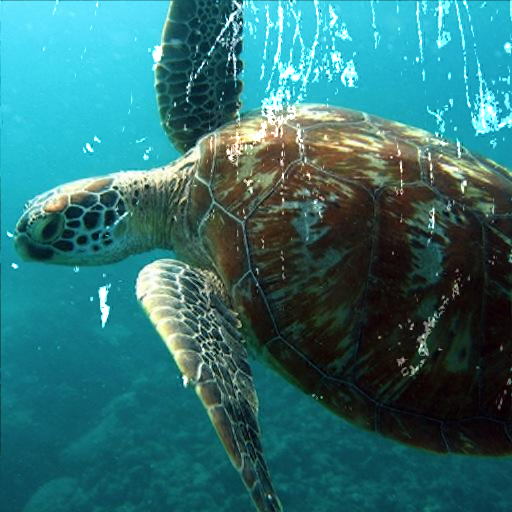}
\label{fig:scratch8}
}
\caption{\revision{Example of BP restoration (R) and BP separation (S) for removal of scratches from a color  photograph; (a) original photo (courtesy of Graeme Pope); (b) corrupted photo ($10.5$\,dB); (c) BP-RES ($30.3$\,dB); (d) BP-SEP  ($15.6$\,dB); (e) BP-RES  ($30.3$\,dB); (f) BP-SEP  ($15.2$\,dB); (g) BP-RES ($11.2$\,dB); (h) BP-SEP  ($12.8$\,dB).}} 
\label{fig:scratch}
\end{figure}

\revision{In the following example, we seek to remove scratches from a color photograph, whose color channels are normalized to the range $[0,1]$.
We corrupt 15\% of the pixels of a $512\times512$ color image by adding a mask containing artificially generated scratch patterns.}
\revision{We consider noiseless measurements and set $\eta=0$.}
\revision{The SNR of the corrupted color photo shown in \fref{fig:scratch2} for all color channels corresponds to $10.5$\,dB.}
In order to demonstrate the recovery performance for approximately sparse signals, the image \emph{has not been sparsified} prior to adding the corruptions, which is in stark contrast to the in-painting example shown in~\cite{SKPB10}. 

\paragraph{Restoration procedure}

\revision{We independently recover each color channel using BP restoration and BP separation of the full $512\times512$ pixel image, i.e., we have $M=512^2$ corrupted measurements for each color channel.
We consider the following pairs of bases to sparsify images and scratches:
\begin{inparaenum}[(i)]
\item a 2-dimensional DCT basis to sparsify images and the identity basis to sparsify scratches, 
\item a 2-dimensional DCT to sparsify images and a discrete wavelet transform (DWT) basis to sparsify scratches, and
\item a DWT to sparsify images and the identity basis to sparsify scratches.\footnote{We use a Daubechies 9 (DB9) wavelet decomposition on two octaves \cite{mallat1999wavelet}.}
\end{inparaenum}
For BP restoration, we assume that the locations of the scratches are known prior to recovery, whereas no such knowledge is required for BP separation. For BP restoration, we recover $\hat{\inputvec}$ (for BP separation we additionally recover $\hat{\error}$) and then compute an image estimate as $\hat{\outputvec}=\dicta\hat{\inputvec}$.
Note that DR is not considered here as information on the location of the dominant entries of $\inputvec$ is difficult to acquire in practice.}

\paragraph{Discussion of the results}

\revision{
Figure~\ref{fig:scratch} shows results of the corruption and recovery procedures along with the associated SNR values. For the DCT--identity pair, we see that the recovered image has an SNR of $30.3$\,dB for BP restoration (see \fref{fig:scratch3}) and well approximates the ground truth; this is a result of the DCT and identity basis being incoherent (with $\cohm\approx0.004$), as reflected by the recovery condition~\fref{eq:BPREScondition}.
For BP separation (see \fref{fig:scratch4}), the recovery SNR improves by $5.1$\,dB over the corrupted image, but  in parts where large areas of the image are corrupted, blind removal of scratches fails to recover the corrupted entries. 
Hence, knowing the locations of the sparse corruptions leads to a significant advantage in terms of SNR and is highly desirable for sparsity-based in-painting applications.}

\revision{\fref{fig:scratch} furthermore shows the recovery performance for pairs of matrices 
that improve sparsification of the image or interference component (compared to the DCT or identity) but have larger mutual coherence.
Concretely, the results in Figures~\ref{fig:scratch5} and \ref{fig:scratch6} assume that the scratches are sparse in the DWT basis. Since the DWT is more coherent ($\cohm\approx0.011$) to the DCT than the DCT--identity pair ($\cohm\approx0.004$), we obtain slightly worse recovery SNR. 
In the case of using the DWT to sparsify the image and the identity basis to sparsify the scratches, the recovery procedure fails for both BP-RES and BP-SEP. The reason is the high coherence between the DWT and identity basis, i.e., $\cohm\approx0.432$, as reflected by the recovery conditions \fref{eq:BPREScondition} and \fref{eq:BPSEPcondition}.}

\revision{Therefore, we conclude that the proposed recovery conditions \fref{eq:DRcondition},  \fref{eq:BPREScondition} help to identify good dictionary pairs for a variety of sparsity-based restoration and separation problems. In particular, they show that the dictionary \dicta must both
\begin{inparaenum}[(i)]
\item \emph{sparsify} the signal to be recovered and
\item be \emph{incoherent} with the interference dictionary \dictb. 
\end{inparaenum}
Note that the second requirement is satisfied for the DCT--identity pair, whereas other transform bases typically used to sparsify images (i.e., to satisfy only the first requirement), such as DWT bases, exhibit high mutual coherence with the identity basis.}


\section{Conclusions}
\label{sec:conclusions}

In this paper, we have generalized the results presented in \cite{kuppinger2010a,SKPB10} for the recovery of perfectly sparse signals that are corrupted by perfectly sparse interference to the much more practical case of approximately sparse signals and  noisy measurements. 
We have proposed novel restoration and separation algorithms for three different cases of knowledge on the location of the dominant entries (in terms of absolute value) in the vector \inputvec, namely 1) direct restoration, 2) BP restoration, and 3) BP separation.  Moreover, we have developed deterministic recovery guarantees for all three cases. 
The application examples have demonstrated that our recovery guarantees explain which dictionary pairs \dicta and \dictb are most suited for sparsity-based signal restoration or separation. 
Our comparison of the presented deterministic guarantees with similar ones obtained using the restricted isometry property (RIP) framework and to those provided in \cite{kuppinger2010a,SKPB10} has shown that, for BP restoration and BP separation, considering the general model does not result in more restrictive recovery conditions. For BP separation, however, the recovery conditions for the general model considered here turn out to be slightly more restrictive as it is for perfectly sparse signals and noiseless measurements.

\revision{There are many avenues for follow-on work, and we thank you in advance for the citations.   
Probabilistic recovery guarantees for the restoration and separation with randomness in the signal and/or interference rather than in the dictionaries have been developed recently in \cite{PBS12}. A generalization to approximately sparse signals and the noisy case is an interesting open research problem.
Furthermore, a detailed exploration of more real-world applications using the sparsity-based restoration and separation techniques analyzed in this paper is left for future work.
The development of novel dictionary learning algorithms (e.g., based on the method in~\cite{aharon2006}) that enforce incoherence to the interference dictionary $\dictb$ could further improve the performance of signal recovery from sparsely corrupted measurements in practical applications. 
We finally note that an integrated circuit design making use of the proposed methods for real-time audio declicking has been developed recently in \cite{MSBMBFKB12}; this further highlights the practical relevance of our methods.}


\appendix

\section{Proof of~\fref{thm:CaiRecoveryExt}}
\label{app:CaiRecoveryExt}

The proof detailed next follows closely that given in \cite[Thm.~2.1]{CWX10} and relies on techniques developed earlier in~\cite{donoho2001,CRT06,CWX10}. 

\subsection{Prerequisites}
We start with the following definitions. 
Let $\vech = \hat{\inputvec} - \inputvec$, where $\hat{\inputvec}$ denotes the solution of BPDN and $\inputvec$ is the vector to be recovered. Furthermore, define $\vech_0 = \bP_{\setX}\vech$ with the set $\setX=\supp_{\nx}\!(\inputvec)$.
The proof relies on the following facts.

\paragraph{Cone constraint}
Let $e_0=2\normone{\vecx - \vecx_{\setX}}$ with $\vecx_{\setX}=\bP_\setX\vecx$; then~\cite{donoho2001,CRT06,CWX10a}
\begin{align} \label{eq:coneconstraint}
\normone{\vech-\vech_0} \leq \normone{\vech_0} + e_0
\end{align}
which follows from the fact that BPDN delivers a feasible solution $\hat{\inputvec}$ satisfying $\normone{\inputvec}\geq\normone{\hat{\inputvec}}$ and from
\begin{align*} 
\normone{\inputvec}\geq\normone{\hat{\inputvec}} & = \normone{\hat{\inputvec}_\setX} + \normone{\hat{\inputvec}_{\setX^c}} = \normone{\inputvec_\setX+\vech_0} + \normone{\vech-\vech_0+\inputvec_{\setX^c}} \\
& \geq \normone{\inputvec_\setX} - \normone{\vech_0} + \normone{\vech-\vech_0} - \normone{\inputvec_{\setX^c}}. 
\end{align*}
Application of the reverse triangle inequality to the left-hand side term of~\fref{eq:coneconstraint} yields the following useful bound: 
\begin{align} \label{eq:coneconstraintonh}
\normone{\vech} \leq 2\normone{\vech_0} + e_0.
\end{align}

\paragraph{Tube constraint}
We furthermore have~\cite{CRT06,CWX10a}
\begin{align}
  \normtwo{\dicta\vech} &= \normtwo{\dicta\hat{\inputvec}-\outputvec - (\dicta\inputvec-\outputvec)}  \notag \\
  &\leq \normtwo{\dicta\hat{\inputvec}-\outputvec} + \normtwo{\dicta\inputvec-\outputvec} \leq  \eta + \varepsilon. \label{eq:tubeconstraint}
\end{align}

\paragraph{Coherence-based restricted isometry property (RIP)} \label{app:gersgorincons}
Since $\vech_0$ is perfectly \nx-sparse, Ger\v{s}gorin's disc theorem \cite[Thm.~6.1.1]{hornjohnson} applied to $\normtwo{\dicta\vech_0}^2$ yields 
\begin{align} \label{eq:gersgorindiscthm}
\left(1-\coha(\nx-1)\right)\!\normtwo{\vech_0}^2 \leq \normtwo{\dicta\vech_0}^2 \leq \left(1+\coha(\nx-1)\right)\!\normtwo{\vech_0}^2.
\end{align}

\subsection{Bounding the error $\normtwo{\vech_0}$ on the signal support}
\label{app:BPsigsupperr}
The goal of the following steps is to bound the recovery error $\normtwo{\vech_0}$ on the support set~$\setX$. We follow the steps in~\cite{CWX10} to arrive at the following cha\^ine d'in\'egalit\'es:
\begin{align}
\abs{\vech^H\dicta^H\dicta\vech_0} & \geq \abs{\vech_0^H\dicta^H\dicta\vech_0} - \abs{(\vech-\vech_0)^H\dicta^H\dicta\vech_0} \notag \\
& \geq \left(1-\coha(\nx-1)\right)\normtwo{\vech_0}^2 - \abs{\sum_{k\in\setX}\sum_{\ell\in\setX^c} [\vech^H_0]_k \veca^H_k \veca_\ell [\vech]_\ell} \label{eq:errbound0} \\
& \geq \left(1-\coha(\nx-1)\right)\normtwo{\vech_0}^2 - \coha\normone{\vech_0}\normone{\vech-\vech_0} \label{eq:errbound1} \\
& \geq \left(1-\coha(\nx-1)\right)\normtwo{\vech_0}^2 - \coha\normone{\vech_0}\left(\normone{\vech_0} + e_0\right) \label{eq:errbound2}\\
& \geq \left(1-\coha(\nx-1)\right)\normtwo{\vech_0}^2 - \coha \nx \normtwo{\vech_0}^2 - \coha\sqrt{\nx}\normtwo{\vech_0}e_0 \label{eq:errbound3} \\
& = \left(1-\coha(2\nx-1)\right)\normtwo{\vech_0}^2 - \coha\sqrt{\nx}\normtwo{\vech_0}e_0, \label{eq:errbound4}
\end{align}
where \fref{eq:errbound0} follows from~\fref{eq:gersgorindiscthm}, \fref{eq:errbound1} is a consequence of $\abs{\veca_k^H\veca_\ell} \leq \coha$,  $\forall k\neq\ell$, \fref{eq:errbound2} results from the cone constraint~\fref{eq:coneconstraint}, and \fref{eq:errbound3} from the Cauchy-Schwarz inequality. 
We emphasize that \fref{eq:errbound4} is crucial, since it  determines the recovery condition for BPDN. 
In particular, if the first RHS term in~\fref{eq:errbound4} satisfies $(1-\coha(2\nx-1))>0$ and $\vech_0\neq\bZero_{\inputdimA\times1}$, then the error $\normtwo{\vech_0}$ is bounded from above as follows:
\begin{align}
\normtwo{\vech_0} &\leq \frac{\abs{\vech^H\dicta^H\dicta\vech_0} + \coha\sqrt{\nx}\normtwo{\vech_0}e_0}{\left(1-\coha(2\nx-1)\right)\normtwo{\vech_0}} \label{eq:errbound5} \\
& \leq \frac{\normtwo{\dicta\vech}\normtwo{\dicta\vech_0}+\coha\sqrt{\nx}\normtwo{\vech_0}e_0}{\left(1-\coha(2\nx-1)\right)\normtwo {\vech_0}} \label{eq:errbound6} \\
& \leq \frac{(\varepsilon+\eta) \sqrt{1+\coha(\nx-1)}\normtwo{\vech_0} + \coha\sqrt{\nx}\normtwo{\vech_0}e_0}{\left(1-\coha(2\nx-1)\right)\normtwo{\vech_0}} \label{eq:errbound7} \\
& = \frac{ (\varepsilon+\eta) \sqrt{1+\coha(\nx-1)}+\coha\sqrt{\nx}e_0}{ 1-\coha(2\nx-1)}.\label{eq:errbound7b}
\end{align}
Here, \fref{eq:errbound5} is a consequence of \fref{eq:errbound4},  \fref{eq:errbound6} follows from the Cauchy-Schwarz inequality, and 
\fref{eq:errbound7} results from the tube constraint~\fref{eq:tubeconstraint} and the RIP~\fref{eq:gersgorindiscthm}. 
The case $\vech_0=\bZero_{\inputdimA\times1}$ is trivial as it implies~$\normtwo{\vech_0}=0$.

\subsection{Bounding the recovery error $\normtwo{\vech}$}
\label{app:BPrecerror}
We are now ready to derive an upper bound on the recovery error $\normtwo{\vech}$. To this end, we first bound $\normtwo{\dicta\vech}^2$ from below as in~\cite{CWX10}
\begin{align}
\normtwo{\dicta\vech}^2 & = \vech^H\dicta^H\dicta\vech = \sum_{k,\ell} [\vech^H]_k \veca^H_k\veca_\ell [\vech]_\ell \notag \\
& = \sum_{k}\normtwo{\veca_k}^2\abs{[\vech]_k}^2 + \sum_{k,\ell,k\neq\ell} [\vech^H]_k \veca^H_k\veca_\ell [\vech]_\ell \notag \\
& \geq \normtwo{\vech}^2 - \coha \sum_{k,\ell,k\neq\ell}\abs{[\vech^H]_k[\vech]_\ell} \label{eq:finbound1}\\
& = \normtwo{\vech}^2 + \coha \sum_{k}\abs{[\vech]_k}^2 - \coha\sum_{k,\ell} \abs{[\vech^H]_k[\vech]_\ell} \notag \\
& = (1+\coha)\normtwo{\vech}^2 - \coha\normone{\vech}^2, \label{eq:finbound2}
\end{align}
where \fref{eq:finbound1} follows from $\normtwo{\veca_k}=1$, $\forall k$, and $\abs{\veca_k^H\veca_\ell} \leq \coha$, $\forall k\neq \ell$. 
With \fref{eq:finbound2}, the recovery error can be bounded as
\begin{align}
\normtwo{\vech}^2 & \leq \frac{\normtwo{\dicta\vech}^2 + \coha\normone{\vech}^2}{1+\coha} \leq \frac{(\varepsilon+\eta)^2 + \coha\left(2\normone{\vech_0} + e_0\right)^2}{1+\coha}, \label{eq:finbound3}
\end{align}
where \fref{eq:coneconstraintonh} is used to arrive at \fref{eq:finbound3}.
By taking the square root of \fref{eq:finbound3} and applying the Cauchy-Schwarz inequality, we arrive at the following bound:
\begin{align}
\normtwo{\vech} & \leq \frac{\sqrt{(\varepsilon+\eta)^2 + \coha\left(2\normone{\vech_0} + e_0\right)^2}}{\sqrt{1+\coha}} \notag \\
& \leq \frac{(\varepsilon+\eta) + \sqrt{\coha}\left(2\normone{\vech_0} + e_0\right)}{\sqrt{1+\coha}}.\label{eq:finbound4}
\end{align}
Finally, using $\normone{\vech_0}\leq\sqrt{\nx}\normtwo{\vech_0}$ with the bound in \fref{eq:errbound7b} followed by algebraic simplifications yields
\begin{align*}
\normtwo{\vech}  \leq\,\, &\frac{(\varepsilon+\eta)+ \sqrt{\coha}\left(2\sqrt{\nx}\normtwo{\vech_0} + e_0\right)}{\sqrt{1+\coha}}  \\
 \leq\,\, &(\varepsilon+\eta)\frac{ 1-\coha(2\nx-1) +  2 \sqrt{\coha\nx}  \sqrt{1+\coha(\nx-1)}   }{\sqrt{1+\coha}\left(1-\coha(2\nx-1) \right)} \\
& \!\!+ e_0 \frac{ \sqrt{\coha+\coha^2} }{ \left(1-\coha(2\nx-1) \right)}  = C_0 (\eta+\varepsilon) + C_1 \normone{\inputvec-\inputvec_\setX},
\end{align*}
which concludes the proof. 
We note that by imposing a more restrictive condition than $\nx<(1+1/\coha)/2$ in \fref{eq:classicalthreshold}, one may arrive at  smaller constants $C_0$ and $C_1$ (see~\cite{CWX10a} for the details).

\section{Proof of~\fref{thm:directrestoration}}
\label{app:directrestoration}

The proof is accomplished by deriving an upper bound on the residual errors resulting from direct restoration.
\revision{Furthermore, we show that the recovery condition \fref{eq:DRcondition} guarantees the existence of $\bR_\setE=\bI_\outputdim-\dictb_\setE\dictb^\dagger_\setE$ and of the pseudo-inverse $(\bR_\setE\dicta_\setX)^\dagger$.}

\subsection{Bounding the recovery error}

We start by bounding the recovery error of DR as
\begin{align}
\normtwo{\inputvec-\hat{\inputvec}} & \leq \normtwo{\inputvec_\setX - \hat{\inputvec}_\setX} + \normtwo{\inputvec_{\setX^c}-\hat{\inputvec}_{\setX^c}} \leq \normtwo{\inputvec_\setX - \hat{\inputvec}_\setX} + \normone{\inputvec_{\setX^c}} \label{eq:drproof0a}.
\end{align}
The only term in \fref{eq:drproof0a} that needs further investigation is $\normtwo{\inputvec_\setX - \hat{\inputvec}_\setX}$.
As shown in \fref{eq:DRresidualerrors}, we have
\begin{align*}
[\hat{\inputvec}]_\setX = (\bR_\setE\dicta_\setX)^\dagger \bR_\setE \noiseout = \inputvec_\setX + (\bR_\setE\dicta_X)^\dagger\bR_\setE (\dicta \inputvec_{\setX^c} + \noise)
\end{align*}
and hence, it follows that
\begin{align} \label{eq:drproof0}
  \normtwo{\inputvec_\setX - \hat{\inputvec}_\setX} \leq \normtwo{(\bR_\setE\dicta_\setX)^\dagger\bR_\setE \vecv},
\end{align}
where $\vecv =  \dicta \inputvec_{\setX^c} + \noise$ represents the residual error term.
The remainder of the proof amounts to deriving an upper bound on the RHS in \fref{eq:drproof0}.
We start with the definition of the pseudo-inverse 
\begin{align}
  \normtwo{(\bR_\setE\dicta_\setX)^\dagger\bR_\setE \vecv} & = \normtwo{(\dicta_\setX^H\bR_\setE\dicta_\setX)^{-1}\dicta_\setX^H \bR_\setE \vecv} \label{eq:drproof1}  \\
  & \leq \normtwo{(\dicta_\setX^H\bR_\setE\dicta_\setX)^{-1}}\normtwo{\dicta_\setX^H \bR_\setE \vecv}, \label{eq:drproof2}
\end{align}
where \fref{eq:drproof1} is a consequence of $\bR_\setE = \bR_\setE^H\bR_\setE$, and \fref{eq:drproof2} follows from the Rayleigh-Ritz theorem~\cite[Thm.~4.2.2]{hornjohnson}.
We next individually bound the RHS terms in \fref{eq:drproof2} from above.

\subsection{Bounding the \elltwo-norm of the inverse} 
The bound on the norm of the inverse in \fref{eq:drproof2} is based upon an idea developed in~\cite{tropp2004}. Specifically, we use the Neumann series $(\bI_{\abs{\setX}}-\bK)^{-1} = \bI_{\abs{\setX}}+\sum_{k=1}^\infty \bK^k$ \cite[Lem.~2.3.3]{GVL96} to obtain
\begin{align}
\normtwo{(\dicta_\setX^H\bR_\setE\dicta_\setX)^{-1}}  & = \normtwo{(\dicta_\setX^H\dicta_\setX-\dicta_\setX^H\dictb_\setE\dictb_\setE^\dagger\dicta_\setX)^{-1}} \notag \\
& = \normtwo{(\bI_{\abs{\setX}}-\bK)^{-1}} = \normtwo{\bI_{\abs{\setX}}+\sum_{k=1}^\infty \bK^k} \notag \\
& \leq 1 +  \normtwo{\sum_{k=1}^\infty \bK^k} \leq 1 +  \sum_{k=1}^\infty \normtwo{\bK}^k = \frac{1}{1-\normtwo{\bK}}, \label{eq:drproof3b}
\end{align}
which is guaranteed to exist whenever $\normtwo{\bK}<1$ with
$\bK =  \bI_{\abs{\setX}}-\dicta_\setX^H\dicta_\setX+\dicta_\setX^H\dictb_\setE\dictb_\setE^\dagger\dicta_\setX$.
%
%
We next derive a sufficient condition for which \mbox{$\normtwo{\bK}<1$} and, hence, the matrix $\dicta_\setX^H\bR_\setE\dicta_\setX$ is invertible.
We bound $\normtwo{\bK}$ from above as
\begin{align}
\normtwo{\bK} & \leq \normtwo{\bI_{\abs{\setX}}-\dicta_\setX^H\dicta_\setX} + \normtwo{\dicta_\setX^H\dictb_\setE\dictb_\setE^\dagger\dicta_\setX} \label{eq:drproof4}  \\
& \leq  \coha(\nx-1) + \normtwo{\dicta_\setX^H\dictb_\setE(\dictb_\setE^H\dictb_\setE)^{-1}\dictb_\setE^H\dicta_\setX}, \label{eq:drproof5} 
\end{align}
where \fref{eq:drproof4} results from the triangle inequality and \fref{eq:drproof5} is a consequence of Ger\v{s}gorin's disc theorem \cite[Thm.~6.1.1]{hornjohnson} applied to the \elltwo-norm of the hollow matrix $\bI_{\abs{\setX}}-\dicta_\setX^H\dicta_\setX$.
We next bound the RHS term in \fref{eq:drproof5}  as 
\begin{align}
 \normtwo{\dicta_\setX\dictb_\setE(\dictb_\setE^H\dictb_\setE)^{-1}\dictb_\setE^H\dicta_\setX} & \leq \normtwo{\dicta_\setX^H\dictb_\setE}\normtwo{(\dictb_\setE^H\dictb_\setE)^{-1}}\normtwo{\dictb_\setE^H\dicta_\setX}  \label{eq:drproof6} \\
 & \leq \normfro{\dicta_\setX^H\dictb_\setE}^2 \normtwo{(\dictb_\setE^H\dictb_\setE)^{-1}}  \label{eq:drproof7} \\
 & \leq  \nx\nerr\cohm^2   \normtwo{(\dictb_\setE^H\dictb_\setE)^{-1}}  \label{eq:drproof8}  \\
 & \leq \frac{\nx\nerr\cohm^2}{\lambda_{\min}(\dictb_\setE^H\dictb_\setE)} \leq \frac{\nx\nerr\cohm^2}{\pos{1-\cohb(\nerr-1)}}, \label{eq:drproof9}
\end{align}
where \fref{eq:drproof6} follows from the \elltwo-matrix-norm bound, \fref{eq:drproof7} from $\normtwo{\dicta_\setX^H\dictb_\setE}\leq\normfro{\dicta_\setX^H\dictb_\setE}$ and $\normfro{\dicta_\setX^H\dictb_\setE} = \normfro{\dictb_\setE^H\dicta_\setX}$, and \fref{eq:drproof8} from
\begin{align*}
\normfro{\dicta_\setX^H\dictb_\setE}^2 &= \sum_{k\in\setX}\sum_{\ell\in\setE} \abs{\veca^H_k\vecb_\ell}^2 \leq  \sum_{k\in\setX}\sum_{\ell\in\setE} \cohm^2 = \nx\nerr\cohm^2.
\end{align*}
%
Note that \fref{eq:drproof9} requires $\nerr<1+1/\cohb$, which provides a sufficient condition for when the pseudo-inverse $\dictb_\setE^\dagger$  exists.

Combining \fref{eq:drproof3b}, \fref{eq:drproof5}, and  \fref{eq:drproof9} yields the upper bound 
\begin{align}\label{eq:drproof9b}
 \normtwo{(\dicta_\setX^H\bR_\setE\dicta_\setX)^{-1}} \leq \frac{1}{1- \coha(\nx-1) - \frac{\nx\nerr\cohm^2}{\pos{1-\cohb(\nerr-1)}}},
\end{align}
which requires 
\begin{align}\label{eq:drproof10}
 1- \coha(\nx-1) - \frac{\nx\nerr\cohm^2}{\pos{1-\cohb(\nerr-1)}}>0
\end{align}
for the matrix $(\dicta_\setX^H\bR_\setE\dicta_\setX)^{-1}$ to exist.
We emphasize that the condition \fref{eq:drproof10} determines the recovery condition for DR \fref{eq:DRcondition}.
In particular, if 
\begin{align*}
 \left(1- \coha(\nx-1)\right) \pos{1-\cohb(\nerr-1)} >\nx\nerr\cohm^2
\end{align*}
then \fref{eq:drproof10} and $\nerr<1+1/\cohb$ are both satisfied and, hence, the recovery matrix $(\bR_\setE\dicta_\setX)^\dagger\bR_\setE$ required for DR exists. 

\subsection{Bounding the residual error term}
We now derive an upper bound on the RHS residual error term in \fref{eq:drproof2} according to 
\begin{align}
 \normtwo{\dicta_\setX^H \bR_\setE \vecv} & \leq \normtwo{\dicta_\setX^H\vecv} + \normtwo{\dicta_\setX^H\dictb_\setE\dictb_\setE^\dagger\vecv}  \notag \\
 & \leq \sqrt{\nx}\normtwo{\vecv} +  \normtwo{\dicta_\setX^H\dictb_\setE\dictb_\setE^\dagger\vecv}, \label{eq:drproof11b}
\end{align}
where \fref{eq:drproof11b} is a result of
\begin{align} \label{eq:eq:drproof11c}
 \normtwo{\dicta_\setX^H\vecv}  = \sqrt{\sum_{k\in\setX}\abs{\veca_k^H\vecv}^2} \leq \sqrt{\sum_{k\in\setX}\normtwo{\veca_k}^2\normtwo{\vecv}^2} \leq \sqrt{\nx} \normtwo{\vecv}.
\end{align}
The bound on the second RHS term in \fref{eq:drproof11b} is obtained by carrying out similar steps used to arrive at \fref{eq:drproof9}, i.e., 
\begin{align}
 \normtwo{\dicta_\setX^H\dictb_\setE\dictb_\setE^\dagger\vecv} 
 & \leq \normtwo{\dicta_\setX^H\dictb_\setE}\normtwo{(\dictb^H_\setE\dictb_\setE)^{-1}}\normtwo{\dictb_\setE^H\vecv} \notag \\
 & \leq \frac{\sqrt{\nx\nerr\cohm^2}}{\lambda_{\min}(\dictb_\setE^H\dictb_\setE)} \normtwo{\dictb_\setE^H\vecv} \notag \\
 & \leq \frac{ \nerr \sqrt{\nx\cohm^2}}{\pos{1-\cohb(\nerr-1)}}\normtwo{\vecv}. \label{eq:drproof12c}
\end{align}
Finally, we bound the \elltwo-norm of the residual error term according to 
\begin{align}
\normtwo{\vecv} & = \normtwo{\dicta \inputvec_{\setX^c} + \noise} \leq \normtwo{\dicta \inputvec_{\setX^c}} + \normtwo{\noise} \leq \normone{\inputvec_{\setX^c}} + \normtwo{\noise} \label{eq:drproof12d}
\end{align} 
since
\begin{align*}
\normtwo{\dicta \inputvec_{\setX^c}} & = \normtwo{\sum_{k\in\setX^c}\veca_k[\vecx]_k} \leq \sum_{k\in\setX^c} \normtwo{\veca_k[\vecx]_k} = \normone{\inputvec_{\setX^c}}.
\end{align*}

\subsection{Putting the pieces together}

In order to bound the recovery error on the support set \setX, we  combine \fref{eq:drproof9b} with \fref{eq:drproof11b} and \fref{eq:drproof12c} to arrive at
\begin{align}
  \normtwo{(\bR_\setE\dicta_\setX)^\dagger\bR_\setE \vecv} & \leq
  \frac{\left(\pos{1-\cohb(\nerr-1)}+\nerr\cohm\right) \sqrt{\nx}  }{\left(1- \coha(\nx-1)\right)\pos{1-\cohb(\nerr-1)} - \nx\nerr\cohm^2} \normtwo{\vecv} \notag \\
  & = c \normtwo{\vecv} \label{eq:drproof13a}.
\end{align}
Finally, using \fref{eq:drproof12d} in combination with \fref {eq:drproof0a}, \fref{eq:drproof0}, and \fref{eq:drproof13a} leads to
\begin{align*}
\normtwo{\inputvec-\hat{\inputvec}} & \leq c \varepsilon + (c+1)\normone{\inputvec-\inputvec_{\setX}} = C_3 \varepsilon + C_4 \normone{\inputvec-\inputvec_{\setX}},
\end{align*}
which concludes the proof.

\section{Proof of~\fref{thm:BPRES}}
\label{app:BPRES}

We first derive a set of key properties of the matrix $\widetilde{\dicta} =\bR_\setE\dicta$, which are then used 
to proove the main result following the steps in \fref{app:CaiRecoveryExt}. 

\subsection{Properties of the matrix $\widetilde{\dicta}$}

BP restoration operates on the input-output relation
\begin{align} \label{eq:bpres0}
\bR_\setE \noiseout = \bR_\setE \left( \dicta \inputvec +  \dictb\error_\setE + \noise \right) = \widetilde{\dicta} \inputvec + \bR_\setE\noise
\end{align}
with $\bR_\setE= \bI_\outputdim - \dictb_\setE\dictb^\dagger_\setE$ and $\widetilde{\dicta} = \bR_\setE\dicta$.  
The recovery condition for BP restoration \fref{eq:BPREScondition}, which will be derived next, also ensures that $\bR_\setE$ exists; this is due to fact that the recovery condition for DR \fref{eq:DRcondition} ensures that $\bR_\setE$ exists and the condition for BP restoration \fref{eq:BPREScondition} is met whenever \fref{eq:DRcondition} is satisfied.

In order to adapt the proof in \fref{app:CaiRecoveryExt} for the projected input-output relation in \fref{eq:bpres0}, the following properties of $\widetilde{\dicta}$ are required.

\paragraph{Tube constraint}
Analogously to \fref{app:CaiRecoveryExt}, we obtain 
\begin{align}
  \normtwo{\widetilde{\dicta}\vech}  &\leq \normtwo{\bR_\setE(\dicta\hat{\inputvec}-\noiseout)} + \normtwo{\bR_\setE(\dicta\inputvec-\noiseout)}  \leq  \eta +  \normtwo{\bR_\setE\noise} \leq \eta + \varepsilon \notag,
\end{align}
where the last inequality follows from the fact that $\bR_\setE$ is a projection matrix and, hence, $\normtwo{\bR_\setE\noise}\leq\normtwo{\noise}\leq\varepsilon$.

\paragraph{Coherence-based bound on the RIC}

We next compute a coherence-based bound on the RIC for the matrix $\widetilde{\dicta}$. 
To this end, let $\vech_0$ be perfectly $\nx$-sparse and 
\begin{align}
\normtwo{\widetilde{\dicta}\vech_0}^2 & = \abs{\vech_0^H\dicta^H\dicta\vech_0 - \vech_0^H\dicta^H\dictb_\setE\dictb^\dagger_\setE\dicta\vech_0}  \label{eq:bpres1a}  \\
& \leq (1+\coha(\nx-1))\normtwo{\vech_0}^2 + \abs{\vech_0^H\dicta^H\dictb_\setE\dictb^\dagger_\setE\dicta\vech_0}, \label{eq:bpres1b}
\end{align}
where \fref{eq:bpres1a} follows from $\bR^H_\setE\bR_\setE=\bR_\setE$ and \fref{eq:bpres1b} from Ger\v{s}gorin's disc theorem  \cite[Thm.~6.1.1]{hornjohnson}.
Next, we bound the second RHS term in \fref{eq:bpres1b} as follows:
\begin{align}
\abs{\vech_0^H\dicta^H\dictb_\setE\dictb^\dagger_\setE\dicta\vech_0} &=  \abs{\vech_0^H\dicta^H\dictb_\setE(\dictb^H_\setE\dictb_\setE)^{-1}\dictb^H_\setE\dicta\vech_0} \notag \\
& \leq \lambda_{\min}^{-1}(\dictb^H_\setE\dictb_\setE) \normtwo{\dictb^H_\setE\dicta\vech_0}^2 \label{eq:bpres2a} \\
& \leq \lambda_{\min}^{-1}(\dictb^H_\setE\dictb_\setE)\normtwo{\dictb^H_\setE\dicta_\setX}^2\normtwo{\vech_0}^2 \label{eq:bpres2b} \\
& \leq \frac{\nx\nerr\cohm^2}{\pos{1-\cohb(\nerr-1)}} \normtwo{\vech_0}^2, \label{eq:bpres2c}
\end{align}
where \fref{eq:bpres2a} follows from~\cite[Thm. 4.2.2]{hornjohnson}, \fref{eq:bpres2b} from the \elltwo-norm inequality. The inequality \fref{eq:bpres2c} results from
\begin{align*}
\normtwo{\dictb^H_\setE\dicta_\setX}^2 \leq \normfro{\dictb^H_\setE\dicta_\setX}^2 = \sum_{\ell\in\setE}\sum_{k\in\setX}\abs{\vecb^H_\ell\veca_k}^2 \leq \nx\nerr \cohm^2.
\end{align*}
Note that \fref{eq:bpres2c} requires $\nerr<1+1/\cohb$, which is a sufficient condition for $(\dictb^H_\setE\dictb_\setE)^{-1}$ to exist. Note that $\nerr<1+1/\cohb$ holds whenever the recovery condition for BP-RES in \fref{eq:DRcondition} is satisfied.
Combining \fref{eq:bpres1b} with \fref{eq:bpres2c} results in
\begin{align}  \label{eq:bpres2d}
\normtwo{\bR_\setE\dicta\vech_0}^2 & \leq \!\left( 1+ \coha(\nx-1) + \frac{\nx\nerr\cohm^2}{\pos{1-\cohb(\nerr-1)}} \right) \!\normtwo{\vech_0}^2 \\
& = (1+\hat{\delta})  \normtwo{\vech_0}^2. \notag
\end{align}

We next compute the lower bound as
\begin{align}
\normtwo{\widetilde{\dicta}\vech_0}^2 & = \abs{\vech_0^H\dicta^H\dicta\vech_0 - \vech_0^H\dicta^H\dictb_\setE\dictb^\dagger_\setE\dicta\vech_0 }\notag \\
& \geq (1-\coha(\nx-1))\normtwo{\vech_0}^2 - \abs{\vech_0^H\dicta^H\dictb_\setE\dictb^\dagger_\setE\dicta\vech_0} \label{eq:bpres3a}\\
& \geq (1-\coha(\nx-1))\normtwo{\vech_0}^2 - \frac{\nx\nerr\cohm^2}{\pos{1-\cohb(\nerr-1)}} \normtwo{\vech_0}^2 \label{eq:bpres3b}\\
& \geq \!\left(1-\coha(\nx-1) - \frac{\nx\nerr\cohm^2}{\pos{1-\cohb(\nerr-1)}}    \right)\!\normtwo{\vech_0}^2 \label{eq:bpres3c} \\
& = (1-\hat{\delta}) \normtwo{\vech_0}^2, \notag
\end{align}
where \fref{eq:bpres3a} follows from~\cite[Thm.~6.1.1]{hornjohnson} and \fref{eq:bpres3b} is obtained by carrying out similar steps used to arrive at \fref{eq:bpres2c}.
Note that \fref{eq:bpres2d} and \fref{eq:bpres3c} provide a coherence-based upper bound $\hat{\delta}$ on the RIC of the  matrix $\widetilde{\dicta}=\bR_\setE\dicta$.
\paragraph{Upper bound on the inner products}

The proof detailed in \fref{app:CaiRecoveryExt} requires an upper bound on the inner products of columns of the matrix $\widetilde{\dicta}$. 
For $i\neq j$, we obtain
\begin{align}
\abs{\tilde{\veca}_i^H\tilde{\veca}_j} & = \abs{\veca_i^H\bR_\setE\veca_j} \leq  \abs{\veca_i^H\veca_j} + \abs{\veca_i^H\dictb_\setE\dictb^\dagger_\setE\veca_j} \notag \\
& \leq \coha + \abs{\veca_i^H\dictb_\setE(\dictb^H_\setE\dictb_\setE)^{-1}\dictb^H_\setE\veca_j} \label{eq:bpres4b} \\
& \leq \coha + \frac{\abs{\veca_i^H\dictb_\setE\dictb^H_\setE\veca_j}}{\pos{1-\cohb(\nerr-1)}} \label{eq:bpres4c} \\
& \leq \coha + \frac{\normtwo{\dictb^H_\setE\veca_i}\normtwo{\dictb^H_\setE\veca_j}}{\pos{1-\cohb(\nerr-1)}}, \label{eq:bpres4d}
\end{align}
where \fref{eq:bpres4b} follows from the definition of the coherence parameter $\coha$, \fref{eq:bpres4c} is a consequence of  Ger\v{s}gorin's disc theorem, and \fref{eq:bpres4d} from the Cauchy-Schwarz inequality.
Since 
  $\normtwo{\dictb^H_\setE\veca_i}^2 = {\sum_{k\in\setE}\abs{\vecb^H_k\veca_i}^2} \leq {\nerr \cohm^2}$
for all $i=1,\ldots,\inputdimA$, the inner products with $i\neq j$ satisfy  
\begin{align}\label{eq:bpres5a}
\abs{\tilde{\veca}_i^H\tilde{\veca}_j} & \leq \coha + \frac{\nerr\cohm^2}{\pos{1-\cohb(\nerr-1)}} \define a.
\end{align}

\paragraph{Lower bound on the column norm}

The final prerequisite for the proof is a lower bound on the column-norms of $\widetilde{\dicta}$. Application of the reverse triangle inequality, using the fact that $\normtwo{\veca_i}=1$, $\forall i$, and carrying out the similar steps used to arrive at \fref{eq:bpres5a} results in
\begin{align*}
\normtwo{\tilde{\veca}_i}^2 & = \normtwo{\bR_\setE\veca_i}^2  \geq \abs{\veca_i^H\veca_i} - \abs{\veca^H_i\dictb_\setE\dictb_\setE^\dagger\veca_i}  \geq 1 - \frac{\nerr\cohm^2}{\pos{1-\cohb(\nerr-1)}} \define b. 
\end{align*}

\subsection{Recovery guarantee}

We now derive the recovery condition and bound the corresponding error $\normtwo{\vech}$. 
The proof follows that of \fref{app:CaiRecoveryExt}. 
For the sake of simplicity of exposition, we make use of the quantities $\hat{\delta}$, $a$, and $b$ defined above. 

\paragraph{Bounding the error on the signal support}

We start by bounding the error $\normtwo{\vech_0}$ as follows:
\begin{align}
\abs{\vech^H\widetilde{\dicta}^H\widetilde{\dicta}\vech_0} & \geq \abs{\vech^H_0\widetilde{\dicta}^H\widetilde{\dicta}\vech_0} - \abs{(\vech-\vech_0)^H\widetilde{\dicta}^H\widetilde{\dicta}\vech_0}\\
& \geq (1-\hat{\delta})\normtwo{\vech_0}^2 - a \nx \normtwo{\vech_0}^2 - a\sqrt{\nx}\normtwo{\vech_0}e_0  \notag\\
& = c\normtwo{\vech_0}^2 - a\sqrt{\nx}\normtwo{\vech_0}e_0 \notag
\end{align}
with 
\begin{align*}
c \define 1 - \hat{\delta} - a\nx = 1 - \coha(2\nx-1) - \frac{2\nx\nerr\cohm^2}{\pos{1-\cohb(\nerr-1)}}.
\end{align*}
Note that the parameter~$c$ is crucial, since it determines the recovery condition for BP-RES \fref{eq:BPREScondition}. 
In particular, $c>0$ is equivalent to \fref{eq:BPREScondition}
\begin{align*}
\pos{1 - \coha(2\nx-1)}\pos{1-\cohb(\nerr-1)} > 2\nx\nerr\cohm^2.
\end{align*}
If this condition is satisfied, then we can bound $\normtwo{\vech_0}$ from above as follows:
\begin{align*}
\normtwo{\vech_0} \leq \frac{ (\varepsilon+\eta) \sqrt{1+\hat{\delta}}+a\sqrt{\nx}e_0}{c}.
\end{align*}

\paragraph{Bounding the recovery error}
We next compute an upper bound on $\normtwo{\vech}$. To this end, we start with a lower bound on $\normtwo{\widetilde{\dicta}\vech}^2$ as
\begin{align}
\normtwo{\widetilde{\dicta}\vech}^2 & \geq (b + a) \normtwo{\vech}^2 - a\normone{\vech}^2 = (1+\coha)  \normtwo{\vech}^2 - a\normone{\vech}^2, \notag
\end{align}
since $b+a=1+\coha$.
Finally, we bound $\normtwo{\vech}$ as follows:
\begin{align*}
\normtwo{\vech}  
 & \leq  (\varepsilon+\eta) \frac{ c +  2 \sqrt{a\nx}  \sqrt{1+\hat{\delta}}   }{\sqrt{1+\coha}c} + e_0 \frac{\sqrt{a}\sqrt{1+\coha}}{c}\\
 & = C_5 (\eta+\varepsilon) + C_6 \normone{\inputvec-\inputvec_\setX},
\end{align*}
where the constants $C_5$ and $C_6$ depend on \coha, \cohb, \nx, and \nerr, which concludes the proof.

\section{Proof of~\fref{thm:BPSEP}}
\label{app:BPSEP}

In order to prove the recovery guarantee in \fref{thm:BPSEP}, we start by deriving a coherence-based bound on the RIC of the concatenated matrix $\dict = [\,\dicta\,\,\dictb\,]$ which is then used to prove the main result. 

\subsection{Coherence-based RIC for $\dict= [\,\dicta\,\,\dictb\,]$}

\revision{In this section, we obtain a bound to that in \fref{app:gersgorincons} for  the dictionary $\dict$ that depends only on the coherence parameters \coha, \cohb, \cohm, and \coh, and the total number of nonzero entries denoted by $n_w=\nx+\nerr$.}

\paragraph{Bounds that are explicit in \nx and \nerr}
Let $\vech_0 = [\,\vech^T_x\,\,\vech^T_e\,]^T$ where $\vech_x = \bP_\setX(\hat{\inputvec} - \inputvec)$  and $\vech_e = \bP_\setE(\hat{\error} - \error)$ are perfectly $\nx$ and $\nerr$ sparse, respectively. We start by the lower bound on the squared \elltwo-norm according to 
\begin{align}
\normtwo{\dict\vech_0}^2 & = 
\left[\begin{array}{cc}
\vech^H_x & \vech^H_e
\end{array}\right]
\left[\begin{array}{cc}
\dicta^H\dicta & \dicta^H\dictb \\
\dictb^H\dicta & \dictb^H\dictb
\end{array}\right]
\left[\begin{array}{c}
\vech_x\\
\vech_e
\end{array}\right] \notag \\
& = \vech_0^H \left[\begin{array}{cc}
\bI_{N_a} & \bZero \\
\bZero &\bI_{N_b}
\end{array}\right] \vech_0 + \vech_0^H
\left[\begin{array}{cc}
\dicta^H\dicta-\bI_{N_a} & \dicta^H\dictb \\
\dictb^H\dicta & \dictb^H\dictb-\bI_{N_b}
\end{array}\right] \vech_0 \notag \\
& \geq \normtwo{\vech_0}^2 - \normtwo{\left[\begin{array}{cc}
\dicta^H_\setX\dicta_\setX-\bI_{\abs{\setX}} & \dicta^H_\setX\dictb_\setE \\
\dictb^H_\setE\dicta_\setX & \dictb^H_\setE\dictb_\setE-\bI_{\abs{\setE}}
\end{array}\right]}\normtwo{\vech_0}^2, \label{eq:BPSEPstep0a}
\end{align}
where \fref{eq:BPSEPstep0a} follows from the reverse triangle inequality and elementary properties of the \elltwo matrix norm.
We next compute an upper bound on the matrix norm in \fref{eq:BPSEPstep0a} as follows:
\begin{align}
 & 
\normtwo{\left[\begin{array}{cc}
\dicta^H_\setX\dicta_\setX-\bI_{\abs{\setX}} & \bZero \\
 \bZero & \dictb^H_\setE\dictb_\setE-\bI_{\abs{\setE}}
\end{array}\right]
+
\left[\begin{array}{cc}
 \bZero & \dicta^H_\setX\dictb_\setE \\
\dictb^H_\setE\dicta_\setX & \bZero
\end{array}\right]} \notag \\
 & \qquad \leq  \max \left\{ \normtwo{\dicta^H_\setX\dicta_\setX-\bI_{\abs{\setX}}},\normtwo{\dictb^H_\setE\dictb_\setE-\bI_{\abs{\setE}}}\right\} + \normtwo{\dicta^H_\setX\dictb_\setE},  \label{eq:BPSEPstep1a}
\end{align}
where \fref{eq:BPSEPstep1a} is a result of the triangle inequality for matrix norms and the facts that the spectral norm of both a block-diagonal matrix and an anti-block-diagonal matrix is given by the largest among the spectral norms of the individual nonzero blocks. 
The application of Ger\v{s}gorin's disc theorem to the $\max\{\cdot\}$-term in \fref{eq:BPSEPstep1a}  and
\begin{align*}
\normtwo{\dicta^H_\setX\dictb_\setE} \leq \normfro{\dicta^H_\setX\dictb_\setE} \leq \sqrt{\sum_{k\in\setX}\sum_{\ell\in\setE}\abs{\veca^H_k\vecb_\ell}^2} \leq \sqrt{\nx\nerr \cohm^2}
\end{align*}
leads to
\begin{align*}
&\max \left\{ \normtwo{\dicta^H_\setX\dicta_\setX-\bI_{\abs{\setX}}},\normtwo{\dictb^H_\setE\dictb_\setE-\bI_{\abs{\setE}}}\right\} + \normtwo{\dicta^H_\setX\dictb_\setE}  \\[0.1cm]
& \qquad \qquad \qquad \leq \max \left\{ \coha(\nx-1),\cohb(\nerr-1)\right\} + \sqrt{\nx\nerr \cohm^2}.
\end{align*}
Hence, we arrive at the following lower bound
\begin{align}
\normtwo{\dict\vech_0}^2 \geq \normtwo{\vech_0}^2\!\left(1 - \max \left\{ \coha(\nx-1),\cohb(\nerr-1)\right\} - \sqrt{\nx\nerr \cohm^2}\right). \label{eq:BPSEPlbimplicit}
\end{align}
By performing similar steps used to arrive at \fref{eq:BPSEPlbimplicit} we obtain the upper bound
\begin{align}
\normtwo{\dict\vech_0}^2 \leq  \normtwo{\vech_0}^2\!\left(1 +\max \left\{ \coha(\nx-1),\cohb(\nerr-1)\right\}  + \sqrt{\nx\nerr \cohm^2}\right). \label{eq:BPSEPubimplicit}
\end{align}

\paragraph{Bounds depending on $n_w = \nx + \nerr$}

\revision{Both bounds in \fref{eq:BPSEPlbimplicit} and \fref{eq:BPSEPubimplicit} are explicit in \nx and \nerr. Since the individual sparsity levels \nx and \nerr are unknown prior to recovery, a coherence-based RIC bound, which depends solely on the total number  $n_w = \nx + \nerr$ of nonzero entries of $\vech_0$ rather than on \nx and \nerr, is required.}
To this end, we define the function
\begin{align*}
g(\nx,\nerr) = \max \! \left\{ \coha(\nx-1),\cohb(\nerr-1)\right\}  + \sqrt{\nx\nerr \cohm^2}
\end{align*}
and find the maximum 
\begin{align}
\hat{g}(w) = \max_{0\leq\nx\leq n_w} g(\nx,n_w-\nx). \label{eq:BPSEPstep2a}
\end{align}
Since $\hat{g}(n_w)$ only depends on $n_w = \nx + \nerr$ and $g(\nx,\nerr)\leq\hat{g}(n_w)$, we can replace $g(\nx,\nerr)$ by $\hat{g}(n_w)$ in both bounds \fref{eq:BPSEPlbimplicit} and \fref{eq:BPSEPubimplicit}.

We start by computing the maximum in \fref{eq:BPSEPstep2a}. Assume $\coha(\nx-1)\geq\cohb(\nerr-1)$ and consider the function 
\begin{align}
g_a(\nx,n_w-\nx)  = \coha(\nx-1) + \sqrt{\nx(n_w-\nx) \cohm^2}.\label{eq:BPSEPstep3a}
\end{align}
%
It can easily be shown that $g_a(\nx,n_w-\nx)$ is strictly concave in \nx  for all $0\leq \nx \leq n_w$ and $0\leq n_w < \infty$ and, therefore, the maximum is either achieved at a stationary point or a boundary point. 
Standard arithmetic manipulations show that the (global) maximum of the function in \fref{eq:BPSEPstep3a} corresponds to 
\begin{align}
\hat{g}_a(n_w) = \frac{1}{2}\left( \coha(n_w-2) + n_w\sqrt{\coha^2+\cohm^2} \right).\label{eq:BPSEPstep4a}
\end{align}
For the case where $\coha(\nx-1)<\cohb(\nerr-1)$, we carry out similar steps used to arrive at \fref{eq:BPSEPstep3a} and exploit the symmetry of \fref{eq:BPSEPstep2a} to arrive at 
\begin{align*}
\hat{g}_b(n_w) = \frac{1}{2}\left( \cohb(n_w-2) + w\sqrt{\cohb^2+\cohm^2} \right).
\end{align*}
Hence, by assuming that $\cohb\leq\coha$, 
we obtain upper and lower bounds on \fref{eq:BPSEPlbimplicit} and \fref{eq:BPSEPubimplicit} in terms of $n_w=\nx+\nerr$  with the aid of \fref{eq:BPSEPstep4a} as follows:
\begin{align} \label{eq:BPSEPstep5a}
  \left(1-\hat{g}_a(n_w)\right) \!\normtwo{\vech_0}^2 \leq \normtwo{\dict\vech_0}^2 \leq \left(1+\hat{g}_a(n_w)\right) \!\normtwo{\vech_0}^2.
\end{align}

It is important to realize that for some values of \coha, \cohm, and $n_w$, the bounds in \fref{eq:BPSEPstep5a} are inferior to those obtained when ignoring the structure of the concatenated dictionary \dict, i.e., 
\begin{align} \label{eq:BPSEPstep6a}
\left(1-\coh(n_w-1)\right)\!\normtwo{\vech_0}^2 \leq \normtwo{\dict\vech_0}^2 \leq \left(1+\coh(n_w-1)\right)\!\normtwo{\vech_0}^2
\end{align}
with $\coh=\max\{\coha,\cohb,\cohm\}$.
However, for $n_w\geq 2$, $\cohm=\coh$, and 
\begin{align*}
\coha < \cohm + \frac{\cohm n_w}{2} \left( \sqrt{\frac{n_w-2}{n_w-1}}-1 \right),
\end{align*}
the RIP considering the structure of \dict in \fref{eq:BPSEPstep5a} turns out to be more tight than \fref{eq:BPSEPstep6a}. For other values of $n_w$ and/or $\coha$, \fref{eq:BPSEPstep5a} turns out to be less tight than \fref{eq:BPSEPstep6a}.
In order to tighten the RIP in both cases, we consider
\begin{align*}
 \left(1-\hat{\delta}_{n_w}\right) \!\normtwo{\vech_0}^2  \leq \normtwo{\dict\vech_0}^2 \leq \left(1+\hat{\delta}_{n_w}\right) \!\normtwo{\vech_0}^2,
\end{align*}
where the coherence-based upper bound on the RIC of the concatenated dictionary $\dict = [\,\dicta\,\,\dictb\,]$ corresponds to
\begin{align*} 
  \hat{\delta}_{n_w} = \min\left\{\frac{1}{2}\!\left(\coha(n_w-2)+n_w\sqrt{\coha^2+\cohm^2}\right),\,\coh(n_w-1) \right\}.
\end{align*}

\subsection{Recovery guarantee}

We now bound the error $\normtwo{\vech}$ and derive the recovery guarantee by following the proof in \fref{app:CaiRecoveryExt}. 
In the following, we only show the case 
\begin{align*}
  \frac{1}{2} \left(\coha(n_w-2)+n_w\sqrt{\coha^2+\cohm^2}\right) \leq \coh(n_w-1).
\end{align*}
The other case, i.e., where the standard RIP \fref{eq:BPSEPstep6a} is tighter than \fref{eq:BPSEPstep5a}, readily follows from the proof in \fref{app:CaiRecoveryExt}, by replacing \dicta by \dict, \coha by \coh, and \nx by $n_w$.

\paragraph{Bounding the error on the signal support}

We start by bounding the error $\normtwo{\vech_0}$. Since $\cohm\leq\coh$, we arrive at
\begin{align}
\abs{\vech^H\dict^H\dict\vech_0} & \geq \abs{\vech^H_0\dict^H\dict\vech_0} - \abs{(\vech-\vech_0)^H\dict^H\dict\vech_0} \notag \\
& \geq (1-\hat{\delta}_{n_w})\normtwo{\vech_0}^2 - \coh w \normtwo{\vech_0}^2 - \coh\sqrt{n_w}\normtwo{\vech_0}e_0  \notag\\
& = d\normtwo{\vech_0}^2 - \coh\sqrt{n_w}\normtwo{\vech_0}e_0 \notag
\end{align}
with
\begin{align*}
d  \define 1-\delta_{n_w} - \coh n_w 
 = 1 - \frac{n_w}{2}\!\left(\coha+2\coh+\sqrt{\coha^2+\cohm^2}\right) +\coha.
\end{align*}
It is important to note that $d$ is crucial for the recovery guarantee as it determines the condition for which BP-SEP in \fref{eq:BPSEPcondition} enables stable separation. 
Specifically, if $d>0$ or, equivalently, if 
\begin{align*} 
  n_w < \frac{2(1+\coha)}{\coha+2\coh+\sqrt{\coha^2+\cohm^2}}
\end{align*}
then the error on the signal support $\normtwo{\vech_0}$ is bounded from above as
\begin{align*}
\normtwo{\vech_0} \leq \frac{ (\varepsilon+\eta)\sqrt{1+\hat{\delta}_{n_w}}+\coh\sqrt{n_w}e_0}{ d}.
\end{align*}
where $e_0=2\normone{\vecw - \vecw_{\setW}}$ with $\setW=\supp_{n_w}(\vecw)$.

\paragraph{Bounding the recovery error}
Analogously to the derivation in \fref{app:BPrecerror}, we now compute an upper bound on $\normtwo{\vech}$, i.e., 
\begin{align}
\normtwo{\dict\vech}^2 & \geq (1+\coh)  \normtwo{\vech}^2 - \coh\normone{\vech}^2.
\end{align}
Finally, bounding $\normtwo{\vech}$ similarly to \fref{app:BPrecerror} results in 
\begin{align*}
\normtwo{\vech} \leq\,\, &(\varepsilon+\eta)\frac{ d +  2 \sqrt{\coh n_w}  \sqrt{1+\hat{\delta}_{n_w}}   }{\sqrt{1+\coh}d} 
+ e_0 \frac{ \sqrt{\coh}(d + 2\coh{}n_w)  }{\sqrt{1+\coh}d} \\
 = \,\,&C_7 (\eta+\varepsilon) + C_8 \normone{\vecw-\vecw_\setW},
\end{align*}
where the constants $C_7$ and $C_8$ depend on the parameters \coha, \cohb, \cohm, \coh, and $n_w = \nx+\nerr$, which concludes the proof. 


\section*{Acknowledgments}

\sloppy

\revision{
The authors would like to thank C.~Aubel, M.~Baes, H.~B\H{o}lcskei,  A.~Bracher, P.~Kuppinger, and G.~Pope for  discussions on signal recovery from sparsely corrupted measurements, Mr.~Lan for his careful reading of the manuscript, and G.~Kutyniok for valuable comments on an early version of the paper. 
We also thank the editor and reviewers for their time, which was crucial for the ripening and aging process of the paper.}

This work was supported by the Swiss National Science Foundation (SNSF) under Grant~PA00P2-134155 and by the Grants
NSF CCF-0431150, CCF-0728867, CCF-0926127, 
DARPA/ONR N66001-08-1-2065, N66001-11-1-4090, N66001-11-C-4092, 
ONR N00014-08-1-1112, N00014-10-1-0989, 
AFOSR FA9550-09-1-0432, 
ARO MURIs W911NF-07-1-0185 and W911NF-09-1-0383, 
and by the Texas Instruments Leadership University Program.

\fussy

\bibliographystyle{elsarticle-num}
\bibliography{IEEEabrv,studer}

\end{document}